\documentclass{article}
\usepackage[onecolumn]{emulateapj}
\usepackage{epsfig}

\newcommand{\tf}{t_{\rm fwhm}}
\newcommand{\tF}{t_F}
\newcommand{\ts}{t_\sigma}
\newcommand{\tst}{t_{\sigma n}}

\begin{document}
\title{Event Rate and Einstein Time Evaluation in Pixel Microlensing}
\author{Edward A. Baltz}
\affil{Department of Physics, University of California, Berkeley, CA 94720}
\and
\author{Joseph Silk}
\affil{Departments of Physics and Astronomy and Center for Particle
Astrophysics, University of California, Berkeley, CA 94720}

\begin{abstract}
It has been shown that a flux--weighted full width at half maximum timescale of
a microlensing event can be used in an unbiased estimator of the optical depth.
For the first time, this allows a physical parameter to be easily estimated
from pixel microlensing data.  We derive analytic expressions for the observed
rate of pixel lensing events as a function of the full width at half maximum
timescale.  This contrasts work in the literature which express rates in terms
of an ``event duration'' or Einstein time, which require knowledge of the
magnification, which is difficult to determine in a pixel event.  The full
width at half maximum is the most directly measured timescale.  We apply these
results to possible pixel lensing surveys, using HST for M87, and CFHT for M31.
We predict M87 microlensing rates for the HST Advanced Camera and for NGST, and
demonstrate that one will be able to probe the stellar IMF.  Next, we describe
a new method by which a crude measurement of the magnification can be made in
the regime of magnifications $A\sim10-100$.  This in turn gives a crude
measurement of the Einstein time.  This program requires good photometry and
sampling in the low magnification tails of an event, but is feasible with
today's technology.
\end{abstract}


\section{Introduction}

The nature of the dark matter in the universe is unknown.  The only direct
detections to date have been of massive baryonic halo objects, MACHOs
(Paczy\'{n}ski 1986), via gravitational microlensing of LMC and SMC stars.
The implications of the microlensing experiments are controversial, since of
the $\sim 20$ events detected, the 3 events with distance determinations are
attributable to star-star lensing (2, and the only SMC events, in the SMC, 1 in
the LMC).  Since the SMC is very extended along the line of sight, one must
wait for more data to draw any definitive conclusions about MACHO location.  In
the case of the LMC, the distribution of timescales for the events suggests
that we should be seeing some halo events. If this is the case, up to half of
the halo dark matter consists of MACHOs of mean mass $\sim 0.5 \rm M_\odot.$

To improve the situation, one must measure many more microlensing events.  This
can be accomplished by constructing much larger cameras and using large
telescopes. However there is a complementary strategy that allows us to
optimize the microlensing yields and that can potentially increase by a factor
of two or more the number of detectable MACHO events. This is the technique of
pixel microlensing (Crotts 1992; Baillon et al. 1993).

Pixel microlensing works as follows.  Individual stars are not resolved.
Nevertheless a microlensing event is detectable as a pixel brightening.  One
can observe with many stars per pixel and thereby allow pixel microlensing to
provide a powerful means of accumulating events.  In fact, pixel microlensing
allows us to use gravitational microlensing to learn about galactic halos even
when individual stars are not resolved.  The high magnifications required have
made it difficult to determine the physical parameters of the source--lens
system.  For example, by observing high magnification events it has not
hitherto been possible to extract such parameters of the source--lens system as
the optical depth.  Recently, it has been shown (Gondolo 1999) that using the
flux--weighted full width at half maximum of the pixel microlensing lightcurve,
one can extract the optical depth.  This motivates us to further explore the
theory of lensing in terms of this and related timescales.

We first define some standard notation.  The source star being lensed lies a
distance $L$ from the observer, and the lens itself lies at a distance $xL$,
thus $0\le x\le 1$.  The Einstein radius is the radius of the Einstein ring of
a massive object of mass $M$.  The Einstein radius and its projected angular
size are given by
\begin{equation}
R_E(x)=\sqrt{\frac{4GMLx(1-x)}{c^2}},\hspace{0.3in}\theta_E(x)=
\sqrt{\frac{4GM(1-x)}{Lc^2x}}.
\end{equation}
Considering the Einstein radius as a function of $x$, a tube from the source to
the observer is traced out.  It is widest at $x=1/2$, and its width goes to
zero as $\sqrt{x}$ or $\sqrt{1-x}$ as $x$ goes to zero or one, respectively.
This is the so-called ``microlensing tube''.  A lens that passes inside its own
microlensing tube produces a lensing event.

The Einstein time is the time that a lens would take to cross the full Einstein
disk if it were to pass through the center,
\begin{equation}
t_E=\frac{2R_E(x)}{v},
\end{equation}
where $v$ is the lens velocity transverse to the line of sight.  The flux
increase in a microlensing event is given by the well known formula (Einstein
1936)
\begin{equation}
\frac{\Delta F(t)}{F}=\delta(u)=f(u^2(t)),\hspace{0.3in}
f(x)=\frac{2+x}{\sqrt{x(4+x)}}-1.
\end{equation}
where $u(t)$ is the perpendicular distance from the the lens to the line of
sight, in units of the Einstein radius $R_E$.  We assume linear motion for the
lens, and a minimum impact parameter $\beta$, giving $u^2(t)=\beta^2+(2\Delta
t/t_E)^2$, with $\Delta t=t-t_0$.  It is useful to note that when $u\ll 1$,
$\delta(u)\approx 1/u$.  This is the so-called high magnification regime.

The optical depth to microlensing is a commonly used quantity.  It is defined
as the number of lenses that are within the microlensing tube for a given
source.  With this definition, the optical depth is clearly given by
\begin{equation}
\tau=\int_0^1\frac{\rho(x)}{M}R_E^2(x)L\,dx=
\frac{4\pi GL^2}{c^2}\int_0^1\rho(x)x(1-x)dx.
\end{equation}
Measurement of the optical depth in a microlensing survey provides a weighted
column density of lenses, and can be used to estimate the total mass in lenses
in galactic halos.

It is straightforward to show (Gondolo 1999) that the full-width at half
maximum of a microlensing event is given by $\tf=t_Ew(\beta)$, with
\begin{equation}
w(\beta)=\sqrt{2f(f(\beta^2))-\beta^2}.
\end{equation}
We find the useful limiting values $w(\beta\ll 1)=\beta\sqrt{3}$ and
$w(\beta\gg 1)=\beta\sqrt{\sqrt{2}-1}$.  For classical lensing, where the
threshold is usually taken to be the Einstein radius, we will need
$w(1)=0.9194$.  The optical depth estimator involves determining the
flux--weighted timescale $\tF=\Delta F_{\rm max}\tf=Ft_E\delta(\beta)w(\beta)$.
In the regime of high magnifications, required by pixel microlensing
experiments, $\beta\ll 1$, giving $\tF\approx Ft_E\sqrt{3}$.  We use flux to
indicate a (dimensionless) number of photons per observation.  This is simply
the number of photons collected in one image of the target.  The excess flux is
then the number of photons in one image above the background.
\section{The Microlensing Rate}

We now would like to derive an expression for the differential microlensing
rate in terms of the timescales that can be easily measured from pixel lensing
data.  We have two choices, $\tf$ and $\tF$.  Recall that it is very difficult
to determine the Einstein time for a pixel event.  For a rate determination
alone, the timescale $\tf$ is usually more useful than the timescale $\tF$.  It
is precisely the full-width at half maximum timescale of an event that
determines if it can be well sampled by a given observing strategy.  The
overall flux comes in to the calculation in determining if a given event can be
detected above some noise level.  If we were to compute the event rates as a
function of $\tF$, the sample strategy and photometry are tied together in a
more complicated way.  We thus advocate quoting the event rate distribution in
$\tf$ rather than $\tF$, even though it is the timescale $\tF$ which is useful
in determining the optical depth from a pixel lensing dataset.  The event rate
is then denoted $d\Gamma/d\tf$.  This is the differential rate of microlensing
events at a given timescale $\tf$.

We modify the rate derivation of Griest (1991) to accommodate the new
timescales.  The differential rate at which lenses enter the microlensing tube
is given by
\begin{equation}
d\Gamma=n(x)f(\vec{v}-\vec{v}_t)v_r\,d^3\vec{v}\,(2R_E\,d\beta)(Ldx)=
\frac{2\rho(x)}{M}f(v)v_r^2R_ELdv_x\,dv_r\,d\beta\,d\alpha\,dx.
\end{equation}
Here $\vec{v}_t$ is the velocity of the microlensing tube in the rest frame of
the galactic halos.  The tube can move because the source and observer are in
general moving in the galactic rest frame.  We define $v_r$ as the transverse
velocity of the lens and $\alpha$ as the angle the transverse velocity makes
with the velocity of the microlensing tube $\vec{v}_t$.  We quote the impact
parameter $\beta$ in units of the Einstein radius.  We will integrate $\beta$
from zero to some maximum value, which means we need the factor of two.  We
assume that the distribution function $f(v)$ is Maxwellian, with dispersion
velocity $v_c$.  The distribution function in the microlensing tube frame is
\begin{equation}
f(v)=\frac{1}{(\pi v_c^2)^{3/2}}\exp\left[-(\vec{v}-\vec{v}_t)^2/
v_c^2\right].
\end{equation}
We see that $(\vec{v}-\vec{v}_t)^2=v_x^2+v_r^2+v_t^2-2v_rv_t\cos\alpha$, and
we can do the integrals in $v_x$ and $\alpha$ easily, giving
\begin{equation}
d\Gamma=\frac{4\rho(x)Lv_c}{M}R_E(x)v^2e^{-v^2}e^{-\eta^2(x)}I_0(2\eta v)dv\,
d\beta\,dx,
\end{equation}
with $v=v_r/v_c$ and $\eta=v_t/v_c$.  Here, $I_0(x)$ is a modified Bessel
function of the first kind.  We further define $\tilde{I}_0(x)=e^{-x}I_0(x)$,
thus $\tilde{I}_0(x\gg 1)\approx 1/\sqrt{2\pi x}$.  Previously, this expression
has been used to get the rate distribution with Einstein time (Alcock et al.
1995).  The variable $\beta$ is trivially integrated to some threshold
$\beta_T$, and we change variables from $v$ to $t_E$.  This gives the
following,
\begin{equation}
\frac{d\Gamma}{dt_E}=\frac{32Lv_c^2\beta_T}{M(v_ct_E)^4}\int^1_0dx\,\rho(x)
R_E^4(x)\exp\left[-\left(\frac{2R_E(x)}{v_ct_E}-\eta(x)\right)^2\right]
\tilde{I}_0\left(\frac{4R_E(x)\eta(x)}{v_ct_E}\right).
\end{equation}
To compute the rate distribution in $\tf$, we instead change variables from
$\beta$ to $\tf$, $2R_Ed\beta=vv_cd\tf/w'(\beta)$.  The transformation is
inverted to obtain $\beta=w^{-1}\left[vv_c\tf/(2R_E)\right]$.  The final
theoretical rate, subject to some threshold to be determined later, is then
\begin{equation}
\frac{d\Gamma}{d\tf}=\frac{2\rho_0Lv_c^2}{M}\int_0^1 dx\,
g(x)\int_0^{\infty}dv\,v^3e^{-[v-\eta(x)]^2}\tilde{I}_0[2\eta(x)v]
\left\{w'\left[w^{-1}\left(\frac{v_c\tf}{2R_E(x)}v\right)\right]
\right\}^{-1}.
\label{eq:rate}
\end{equation}
If we instead had changed variables from $v$ to $\tf$, we would have obtained a
double integral in $\beta$ and $x$, with identical functional form in the limit
of small $\beta$.  We choose to integrate in $v$ and $x$ because it is perhaps
more suggestive to integrate over the velocity distribution of lenses, then
along the line of sight.

We now must account for the fact that not all events are detectable.  The first
difficulty is that there must be some minimum of additional flux detected so
that the event is distinguishable from noise.  The simplest way to do this is
the so-called ``peak threshold''.  This is the prescription that an event must
have some fixed maximum amplification for it to be detectable.  Effectively, we
demand that the signal to noise of the measurement at peak magnification be
higher than some threshold.  We assume that electrons in the CCD are collected
at a rate $S_010^{-0.4m}$, where $m$ is the apparent magnitude in some band,
and we assume an integration time $t_{\rm int}$.  The signal to noise of the
highest magnification sample is thus
\begin{equation}
Q=\delta_{\rm max}\sqrt{S_0t_{\rm int}}10^{-0.4(m-\mu/2)},
\end{equation}
where $m$ is the apparent magnitude of the unresolved lensed star and $\mu$ is
the apparent magnitude of the unlensed galaxy light on the pixel.  We see that
a given signal to noise requires some minimum value of $\delta$, which
corresponds to a maximum impact parameter, $\beta_T$.  This gives an upper
bound on $v$, as we will see.  Fixing all variables except $v$ and $\beta$, we
find that $v$ is monotonic in $\beta$, thus we arrive at an upper bound on $v$.
\begin{equation}
\frac{v_c\tf}{2R_E(x)}v=w(\beta)\rightarrow v_{\rm max}^p=
\frac{2R_E(x)}{v_c\tf}w\left(\beta_T\right).
\end{equation}

To improve the peak threshold trigger, we can use the signal to noise in the
three highest points of the lightcurve.  We will assume that the middle sample
is on the peak, and that the samples are a time ${\cal N}t_{\rm int}$ apart.
The significance of the value ${\cal N}$ is that the observing program is
engaged in on--source integration a fraction $1/{\cal N}$ of the time.  The
side samples are then taken at times ${\cal N}t_{\rm int}$ off of the peak.
The signal to noise as measured this way is given by
\begin{equation}
Q^2=S_0t_{\rm int}10^{-0.8(m-\mu/2)}\left[\delta^2(\beta)+2\delta^2\left(
\sqrt{\beta^2+4{\cal N}^2w^2(\beta)t_{\rm int}^2/\tf^2}\right)\right].
\end{equation}
This expression for $Q^2$ is monotonic in $\beta$, and can be inverted to
determine a maximum impact parameter as before.  This maximum impact parameter
then gives the maximum velocity.

A more sophisticated way to account for the detection of an event is the
so-called ``event threshold'' (Gould 1995).  The signal to noise in one sample
is given by
\begin{equation}
Q_i=\sqrt{S_0t_{\rm int}}\delta_i10^{-0.4(m-\mu/2)}.
\end{equation}
The total signal to noise for the event is thus given by
\begin{equation}
Q^2=\sum Q_i^2=S_0t_{\rm int}10^{-0.8(m-\mu/2)}\sum\delta_i^2,\hspace{0.3in}
\sum\delta^2_i\rightarrow\frac{1}{{\cal N}t_{\rm int}}\int dt\,\delta^2(t).
\end{equation}
In converting the sum to an integral, we need to include the factor ${\cal N}$
for the observation spacing, since the telescope does not integrate
continuously.  This factor is neglected in previous work (Gould 1995),
resulting in overestimates of rates by a factor of three or more.  Following
Gould, we find
\begin{equation}
Q^2=\frac{R_E(x)S_0\pi}{v_c{\cal N}}10^{-0.8(m-\mu/2)}
\frac{\zeta(\beta)}{v\beta}.
\end{equation}
The correction function $\zeta$ can be expressed in terms of complete elliptic
integrals with {\em parameter} $m=4/(4+\beta^2)$.
\begin{equation}
\zeta(\beta)=1-\frac{4\beta}{\pi\sqrt{4+\beta^2}}
\left[\frac{\pi}{4}+(2+\beta^2)K(m)-(4+\beta^2)E(m)\right],
\end{equation}
where $K$ and $E$ are complete elliptic integrals of the first and second
kinds, respectively.  Note that some authors define $K$ and $E$ as functions of
the {\em modulus} $k$, with $m=k^2$.  The limiting behavior of this function is
given by $\zeta(\beta\ll 1)=1$ and $\zeta(\beta\gg 1)=(5/4)\beta^{-6}$.  For
$\beta<0.2$, the leading order expansion is accurate to $1\%$:
$\zeta(\beta)=1-2\beta/\pi[\pi/4-4+\ln(8/\beta)]$.  We find that when fixing
all variables except $v$, $Q^2(v)$ is monotonically decreasing.  Thus if we
impose a threshold signal to noise for detection, $Q_0$, we find that $v_{\rm
max}^e=Q^{-1}(Q_0)$.  As a final point, in the event threshold case $Q^2$ is
also monotonic in $\beta$, all other variables being fixed.  This means that we
can translate the maximum velocity into a maximum impact parameter,
\begin{equation}
\beta_T=w^{-1}\left(\frac{v_c\tf}{2R_E(x)}v_{\rm max}^e\right).
\end{equation}
In practice it is quite difficult to implement the event threshold trigger
because of non-poissonian errors in the background estimation (Gondolo 1998).
Using this formalism is more appropriate for space-based observations, which
have much more easily characterized errors.

We must now consider the effects of the finite sizes of the source stars.  The
microlensing formulas assume that the lens and the source are pointlike, but
this is obviously not the case.  The source star has a non-zero radius.  The
nature of the lens is a priori unknown, but the finite lens radius is
unimportant for most possibilities.  The maximum amplification of an extended
source assumed to be uniform can be determined by integrating the magnification
over the disk of the star.  The result is
\begin{equation}
A_{\rm max}=2\theta_E/\theta_{\rm s}=4\sqrt{\frac{GML}{R_s^2c^2}}
\sqrt{\frac{1-x}{x}},
\label{eq:finite}
\end{equation}
where the $\theta$'s are projected angular sizes of the Einstein ring and the
source star, respectively.  Following Gould (1995), we take the radius of a
star of $M_I=$1.0, 0.5, 0.0,... -3.0 to be $R_s/R_\odot=$ 7, 9, 13, 17, 23, 33,
46, 64, and 90.  Such bright red stars are predominantly giants.  The maximum
amplification of a given star is a monotonically decreasing function of $x$.
In the peak threshold case, we require that an event has a minimum
amplification $A_{\rm min}$, which corresponds to maximum value of $x$ where
$A_{\rm min}=A_{\rm max}$,
\begin{equation}
x_{\rm max}=\left[1+\frac{A_{\rm min}^2}{16}\frac{R_s^2c^2}{GML}\right]^{-1}.
\end{equation}
This is easily accounted for the the rate integral.

In the event threshold case, we could take the approach that there is a minimum
amplification as in the peak threshold case.  However, we choose to take a very
conservative approach.  We demand that for an event to be detectable, the lens
must not pass in front of the disk of the star.  This imposes a minimum value
of the impact parameter $\beta_{\rm min}=\theta_s/\theta_E$.  Fixing all other
variables, this imposes a minimum velocity,
\begin{equation}
v_{\rm min}=\frac{2R_E(x)}{v_c\tf}w\left(\frac{\theta_s}{\theta_E}\right).
\end{equation}
In summary, the peak threshold formalism imposes a maximum value of $x$ and a
maximum value of $v$ on Eq.~\ref{eq:rate}, while the event threshold imposes
both a minimum and maximum value of $v$ on Eq.~\ref{eq:rate}.

To get a better feel for the meaning of these results, we compare the bounds on
the velocity integral for the peak threshold and event threshold triggers in
the regime where we require high magnifications.  In other words, we can assume
that $\beta\ll 1$ and $\delta=1/\beta$.  We will assume that we want $Q\ge
Q_0$.  Thus, in the peak threshold case, we find
\begin{equation}
\beta_T=\frac{\sqrt{S_0t_{\rm int}}}{Q_0}10^{-0.4(m-\mu/2)}\;{\rm and}\;
v_{\rm max}^p=2\sqrt{3}\,\frac{R_E(x)}{v_c\tf}\frac{\sqrt{S_0t_{\rm int}}}{Q_0}
10^{-0.4(m-\mu/2)}.
\end{equation}
Similarly, in the event threshold case we find
\begin{equation}
v_{\rm max}^e=\frac{R_E(x)}{v_c}\sqrt{\frac{S_0}{\tf}}
\sqrt{\frac{\pi}{{\cal N}}}10^{-0.4(m-\mu/2)}\frac{1}{Q_0}.
\end{equation}
For comparison, we form the ratio of these maximum velocities,
\begin{equation}
\frac{v_{\rm max}^e}{v_{\rm max}^p}=
\sqrt{\frac{\pi}{2\sqrt{3}}}\sqrt{\frac{\tf}{{\cal N}t_{\rm int}}}\approx
0.95\sqrt{\frac{\tf}{{\cal N}t_{\rm int}}}.
\end{equation}
In the limit where the event is well sampled, $\tf\gg{\cal N}t_{\rm int}$, it
is clear that the event threshold prescription allows the detection of many
more events because the experiment is sensitive to higher velocity lenses.

Now that we have derived the microlensing rate for a single star with magnitude
$m$ which falls on a pixel of magnitude $\mu$, we need to integrate over the
luminosity function of the pixel in order to get the total rate of pixel
lensing events.  We assume that we know the normalized luminosity function of
the pixel, $\phi(m)$, in units of stars per unit magnitude.  The total
microlensing rate is then simply
\begin{equation}
\frac{d\Gamma}{d\tf}=\int^\infty_{-\infty}dm\,\phi(m)\frac{d\Gamma}{d\tf}(m).
\end{equation}
For clarity, we now write the total microlensing rate in both the peak
threshold and event threshold formalisms,
\begin{eqnarray}
\frac{d\Gamma^p}{d\tf}=&
\mbox{$\displaystyle\frac{2\rho_0Lv_c^2}{M}\int^\infty_{-\infty}dm\,\phi(m)$}&
\int_0^{x_{\rm max}(m)} dx\, g(x)\times \\
&&\int_0^{v^p_{\rm max}(m)}dv\,
v^3e^{-[v-\eta(x)]^2}\tilde{I}_0[2\eta(x)v]
\left\{w'\left[w^{-1}\left(\frac{v_c\tf}{2R_E(x)}v\right)\right]
\right\}^{-1}, \nonumber \\
\frac{d\Gamma^e}{d\tf}=&
\mbox{$\displaystyle\frac{2\rho_0Lv_c^2}{M}\int^\infty_{-\infty}dm\,\phi(m)$}&
\int_0^1dx\,g(x)\times \\
&&\int_{v_{\rm min}(m)}^{v^e_{\rm max}(m)}dv\,
v^3e^{-[v-\eta(x)]^2}\tilde{I}_0[2\eta(x)v]
\left\{w'\left[w^{-1}\left(\frac{v_c\tf}{2R_E(x)}v\right)\right]
\right\}^{-1}. \nonumber
\end{eqnarray}
These expressions are triple integrals.  With a reasonable choice of grid,
logarithmic in $x$ or $1-x$, linear in $v$ and $m$, they can be evaluated to
$1\%$ accuracy without much computational difficulty.

For completeness, we would also like an expression for the differential
microlensing rate in terms of the flux-weighted timescale $\tF$.  We return to
the differential relation for the rate, and we include the dependence on the
luminosity function explicitly,
\begin{equation}
d\Gamma=\frac{4\rho(x)Lv_c}{M}R_E(x)\phi(m)v^2e^{-[v-\eta(x)]^2}
\tilde{I}_0[2\eta(x)v]dv\,d\beta\,dx\,dm.
\end{equation}
We will make a change of variables from $m$ to $\tF$.  We find that
$\tF=S_0t_{\rm int}10^{-0.4m}2R_E(x)/(v_cv)w(\beta)\delta(\beta)$, and
furthermore that $dm=-2.5\,d\tF/(\tF\ln 10)$.  With the peak threshold
triggers, where there is a clear threshold impact parameter $\beta_T$, we can
easily find an expression for the rate,
\begin{equation}
\frac{d\Gamma}{d\tF}=\frac{10}{\ln 10}\frac{\rho_0Lv_c}{Mt_F}
\int_0^{\beta_T}d\beta\int_0^1dx\,g(x)R_E(x)\int_0^\infty dv\,
v^2e^{-[v-\eta(x)]^2}\tilde{I}_0[2\eta(x)v]
\phi\left[-2.5\log\left(\frac{v_c\tF v}
{2R_E(x)w(\beta)\delta(\beta)S_0t_{\rm int}}\right)\right].
\end{equation}
When a high magnification is required, $\beta_T\ll 1$ and
$w(\beta)\delta(\beta)\approx\sqrt{3}$, giving a formula involving one fewer
integral,
\begin{equation}
\frac{d\Gamma}{d\tF}=\frac{10}{\ln 10}\frac{\rho_0Lv_c\beta_T}{Mt_F}
\int_0^1dx\,g(x)R_E(x)\int_0^\infty dv\,
v^2e^{-[v-\eta(x)]^2}\tilde{I}_0[2\eta(x)v]
\phi\left[-2.5\log\left(\frac{v_c\tF v}
{2\sqrt{3}R_E(x)S_0t_{\rm int}}\right)\right].
\end{equation}
As stated before, the timescale $\tF$ can be used in an unbiased estimator of
the optical depth, as shown by Gondolo (1999).  However, it is more difficult
to interpret the experimental meaning of $\tF$, and it is probably more useful
to use the timescale $\tf$ in discussing pixel microlensing rates.  Of course,
the physical meaning of $\tF$ is more clear: in the high magnification regime,
it is just proportional to the absolute flux of the source star times the
Einstein time, with no dependence on the impact parameter.

\section{Measuring The Einstein Time}

The flux--weighted full-width at half maximum timescale of an event can be used
to determine the optical depth.  We are still interested, however, in the
physical parameters of the source--lens system, encoded in the Einstein time.
The difficulty in pixel microlensing is that the shape of the lightcurve does
not easily allow the determination of the Einstein time at high magnifications.
In fact, shape alone is not a good indicator of the Einstein time even with
small magnifications (Gould 1996; Wo\'{z}niak \& Paczy\'{n}ski 1997).  In
microlensing with resolved stars, there is extra information in that the flux
of the source star is measurable.

With the estimator of Gondolo in mind, we have a situation different than that
of previous work.  We can measure the optical depth to microlensing with no
knowledge of the Einstein time.  This already gives us a great deal of
information about the MACHO population in a galaxy.  With this information in
hand, we can attempt to determine the Einstein times for some subset of the
data which has high signal to noise.

In the high magnification regime, we see that $\delta\approx 1/u$, giving the
shape of the lightcurve
\begin{equation}
L(t)=B+\frac{F}{\beta}\left(1+\frac{4\Delta t^2}{\beta^2t_E^2}\right)^{-1/2},
\end{equation}
where $B$ is the background and $F$ is the flux of the unlensed star.  At high
magnifications, we see that even several times $\tf$ off of the peak, the
magnification is well characterized by $\delta\approx 1/u$.  In this regime we
can only measure the quantities $F/\beta$ and $\beta t_E\approx \tf/\sqrt{3}$.

In order to measure the three parameters $F$, $\beta$, and $t_E$ separately, we
need to observe the deviations from $\delta=1/u$.  Gould (1996) has shown that
this is possible with pixel lensing data with signal to noise $Q>50$.  To next
order in $u$, the lightcurve is
\begin{equation}
L(t)=B-F+\frac{F}{\beta}\left(1+\frac{4\Delta t^2}{\beta^2t_E^2}\right)^{-1/2}+
\frac{3}{8}F\beta\left(1+\frac{4\Delta t^2}{\beta^2t_E^2}\right)^{1/2}.
\end{equation}
This form contains both $F/\beta$ and $F\beta$, breaking the degeneracy.
Unfortunately, the $F\beta$ term is quite small, requiring a high signal to
noise to obtain a robust fit, as Gould describes.

We describe a new method for measuring the three degenerate parameters in pixel
lensing data using another timescale that can in principle be directly measured
from pixel microlensing data.  This is the standard deviation (in time) of
$\delta[u(t)]$.  We define this statistical timescale as follows,
\begin{equation}
\ts^2=\frac{\int^\infty_{-\infty}dt\,\Delta t^2\delta(u(t))}
{\int^\infty_{-\infty}dt\,\delta(u(t))}=t_E^2\xi^2(\beta).
\label{eq:ts}
\end{equation}
We can evaluate $\xi$ in terms of complete elliptic integrals,
\begin{equation}
\xi(\beta)=\sqrt{\frac{4+\beta^2}{12}}
\left[\frac{(2+\beta^2)E(m)-\beta^2K(m)}{(2+\beta^2)K(m)-(4+\beta^2)E(m)}
\right]^{1/2},
\end{equation}
where $K(m)$ and $E(m)$ are the complete elliptic integrals of the first and
second kinds as functions of the parameter $m=4/(4+\beta^2)$.  We find the
limiting behavior $\xi(\beta\gg 1)=\beta/2$, $\xi(\beta\ll
1)=\left[3\ln(8/\beta)-6\right]^{-1/2}$, and $\xi(1)$=0.778 for events tangent
to the Einstein ring.

We now see the utility of $\ts$.  At high magnifications, corresponding to
small $\beta$, the functional forms of $\tf$ and $\ts$ are quite different.  If
we form the ratio $\ts/\tf$, the Einstein times cancel, and we are left purely
with a function of $\beta$.  Thus, if we could measure both $\tf$ and $\ts$ for
an event, we could measure $\beta$, and thus the magnification and the Einstein
time.

We find that $\ts\gg\tf$ in the high magnification regime.  In other words, the
tails of the lightcurve are very wide.  This raises the question of how well
$\ts$ can be measured.  It turns out that $\ts$ cannot be measured without
extending to hundreds of times $\tf$ in the monitoring program.  However, we
can modify this timescale to make it useful.  For a given event, we assume that
the lightcurve is measured in the interval $(-n\tf,n'\tf)$.  We choose to
truncate the integrals in Eq. (\ref{eq:ts}) at these multiples of $\tf$, for
each event.  In other words, we measure $\tf$ for an event, then we compute the
second moment, truncating the integrals.  This defines a new timescale, which
we denote the truncated statistical timescale,
\begin{equation}
\tst^2=\frac{\int_{0}^{n\tf}dt\,\Delta t^2\delta(u(t))}
{\int_{0}^{n\tf}dt\,\delta(u(t))}=t_E^2\xi^2_n(\beta).
\label{eq:tst}
\end{equation}
The function $\xi_n$ is given by
\begin{equation}
\xi_n(\beta)=\frac{1}{2\sqrt{3}}
\left[\frac{(x_n^2-2)x_nf_n-x_n^3+\sqrt{4+\beta^2}[(2+\beta^2)E(\phi_n|m)-
\beta^2F(\phi_n|m)]}{x_n(f_n-1)+[(2+\beta^2)F(\phi_n|m)-
(4+\beta^2)E(\phi_n|m)]/\sqrt{4+\beta^2}}\right]^{1/2},
\end{equation}
with $x_n=2nw(\beta)$, $f_n=[(4+\beta^2+x_n^2)/(\beta^2+x_n^2)]^{1/2}$, and
$\phi_n=\tan^{-1}(x_n/\beta)$.  Note that $F$ and $E$ are elliptic integrals of
parameter $m=4/(4+\beta^2)$ as before, and that $\xi_n$ approaches $\xi$ for
large $n$.  For the full event, with data in the range $(-n\tf,n'\tf)$, we can
use the previous formula, and simply duplicate each term in the numerator and
denominator replacing $n$ with $n'$.

The real focus of this method is the following.  In pixel microlensing, it is
easy to measure the width of the peak, $\tf$.  Events with the same $\tf$ have
drastically different magnifications.  Our method attempts to measure the width
of the tails of the event, $\ts$, which breaks the degeneracy seen in $\tf$.
In effect, $\tf$ sets the overall timescale of the event, while $\ts$ splits
the degeneracy in events of a given timescale.  It is the comparison of the two
widths, the scaling of one in terms of the other, which allows us to determine
the physical parameters of the source--lens system.

The truncated statistical timescale offers another piece of information about a
microlensing event.  Because it is only a moment of the lightcurve, we can hope
to be able to measure it with some accuracy.  To illustrate the application of
this method, we form the appropriate ratio, $\tst/\tf$, and plot versus $\beta$
for $n=10,20,\infty$ in figure \ref{fig:tsigma}.  Note that $n=\infty$ is the
idealized case, giving $\ts/\tf$.  For $n=10$, the ratio varies by a factor of
more than three between magnifications of a few and magnifications of a
hundred.  A reasonably good measurement of the ratio can thus give a crude
measurement of the true magnification.

\section{Observational Possibilities}

To illustrate these results we specialize to the case of M87, a giant
elliptical in the Virgo cluster.  We take the distance modulus to Virgo to be
$D=31$.  We use the $I$-band luminosity function of Terndrup et al., measured
for the bulge of the Milky Way.  We interpolate linearly in $(M_I, \log N)$ as
follows: (-3.5, 0.0), (-0.6, 1.0), (0.0, 1.5), (0.5, 1.3), (1.25, 1.5), (1.5,
1.7), (2.5, 2.1), (3.12, 2.65), (4.41, 2.95), (7.35, 3.32), (10.0, 3.9).  At
the distance modulus of 31, these correspond to $m_I$ in the range
$m_I=27.5-41$.

Pixel lensing observations of M87 require HST or the future NGST.  For WFPC2 on
HST, we find that $S_0=4.5\times 10^9$ s$^{-1}$.  As proposed by Gould (1995),
we consider using WFPC2, centering the PC chip on the center of the galaxy.
With a program of dithering, we can achieve a PSF of 1 WFPC2 pixel, which is
0.1\arcsec\ on a side.  For simplicity we assume that we can obtain similar
resolution on the PC chip, taking the PSF to be 4 PC pixels, each
0.0455\arcsec\ on a side.  Since each of the three WFC and the single PC chips
is 800 pixels on a side, we effectively have $2.08\times 10^6$ pixels.  M87 is
at a relatively low declination, which means that in an HST orbit of 96
minutes, we can obtain about $t_{\rm int}=$1800 s of on-target integration.
Notice that this implies that ${\cal N}\ge 3$ for HST observations of M87,
since the telescope can only integrate for one third of each orbit.  This fact
is neglected in previous work (Gould 1995).  We also compute rates for the
Advanced Camera (AC) to be installed on HST in the near future.  This CCD has
4096 $0.05\arcsec$ pixels on a side, and has an efficiency in the $I$-band
approximately seven times as large as WFPC2.  We take $S_0=3.15\times 10^{10}$
s$^{-1}$ for the Advanced camera.  Finally, we compute rates for the proposed
Next Generation Space Telescope.  We assume a CCD that is 8192 $0.03\arcsec$
pixels on a side with comparable efficiency to the Advanced camera, and an
aperture of 7.2 meters, three times as large as the HST.  This gives
$S_0=2.84\times 10^{11}$ s$^{-1}$ for the NGST.

We assume that the Milky way halo is an isothermal sphere with a core of 5 kpc
and a dispersion velocity of 220 km s$^{-1}$.  The density is normalized to be
0.3 GeV cm$^{-3}$ at the solar circle.  We truncate the halo at a radius of 200
kpc.  We assume that M87 has a very similar halo, though ten times as massive
and with a dispersion velocity of 695 km s$^{-1}$.  We take the core and
truncation radii to be 5 kpc and 200 kpc respectively.

We now compare the rates of observable events in several cases.  First, we
investigate the peak threshold trigger.  In this case the rate at which
observations are taken only determines the range of timescales of observable
microlensing events.  The sample rate determines the smallest timescale, and
the total duration of monitoring determines the largest timescale.  The rate
distributions for $M=0.01, 0.1, 1.0$ are plotted for WFPC2, AC, and NGST in
figure \ref{fig:peak}.  We have assumed that $Q_0=7$ is required for detection.
Notice that most observable events are quite short.  We also plot the event
rates for the Advanced Camera with $Q_0=42$, appropriate for attempting to
determine the Einstein time of individual events.

In figure \ref{fig:peak3} we plot event rates using the improved peak threshold
trigger, assuming that we make one observation per day.  For comparison, we
present the rates for the Advanced Camera assuming that we make four equally
spaced observations in one day.  Using the three highest samples in the
lightcurve to define an event, we find significantly improved detection rates
for timescales longer than the spacing of the observations.

We plot the same four sets of rates using the event threshold trigger in figure
\ref{fig:event}.  Again, we notice a significantly enhanced rate of observable
events with timescales longer than the spacing between observations.

Microlensing can also occur when the lens itself is also a star.  The rates for
star--star lensing are computed in much the same way.  We assume that a typical
star has a mass of $0.4 M_\odot$ (Gould, Bahcall, \& Flynn 1996), and that the
stars are in an isothermal distribution, with a core radius of 2 kpc.  The
total mass in stars for M87 is taken to be the same as the total dark mass of
the Milky Way.  It is possible that there are a great number of brown dwarfs in
stellar systems.  We also consider the case where the typical star mass is 0.05
$M_\odot$, indicating that most of the mass in stars is in brown dwarfs.  In
figure \ref{fig:starstar}, we plot these event rates in four cases, all with
the improved peak threshold trigger.

The Andromeda Galaxy, M31, has been the subject of several ground--based pixel
microlensing surveys (Crotts 1992; Crotts \& Tomaney 1996; Tomaney \& Crotts
1996; Baillon et al. 1993; Ansari et al. 1997).  This galaxy is a somewhat
peculiar target for a microlensing survey.  The disk of the galaxy is sharply
inclined, only 12.5$^\circ$ from edge-on.  Because most microlensed stars lie
in the disk, there is a definite bias in the distance to the lensed star.
There will be a noticeably higher optical depth to stars on the far side of the
disk.  We take the M31 halo to be identical to the Milky Way halo previously
described, except that it is twice as massive.  In figure \ref{fig:contour}, we
show contours of constant optical depth to the M31 disk.  The isophotes are
taken from Hodge \$ Kennicutt (1982).  The change in optical depth from the
near to far edge is clearly seen.  We now compute event rates for this galaxy.
We take the distance modulus of $D=24.3$.  This smaller distance modulus allows
us to observe pixel microlensing from a ground based observatory.  The CFHT 3.6
meter telescope on Mauna Kea is ideal for this purpose.  It has a new
wide-field CCD, the CFHT12k, which has 12288$\times$8192 $0.2\arcsec$ pixels,
for a field of view of $41\arcmin\times 27\arcmin$.  We take a PSF of $5\times
5$ ``super--pixels'', making a square arc second, conservative for CFHT.  We
assume that $S_0=5.66\times 10^{10}$, and furthermore that the noise level is
twice the photon counting noise, a level obtained by the AGAPE collaboration
(Ansari et al. 1997).  We use a $V$-band luminosity function appropriate for
the disk of a spiral galaxy (Bahcall \& Soneira 1980).  Furthermore, we
truncate the luminosity function at the bright end when it implies a mean of
one star per super--pixel of the brightness of the pixel.  This is a crude way
to remove the events involving resolved stars.  As before, we compute rates for
the peak and improved peak threshold triggers when the sampling rate is once
per day.  We will assume 3600 second exposures.  We center the CCD on M31, and
align its long axis along the major axis of the galaxy.  In most cases we take
$Q_0=7$, though we also compute the rates for $Q_0=42$ with the improved peak
threshold trigger.  In addition, there are a large number of resolved stars in
this program.  We compute the rates of resolved microlensing events, assuming
that a flux increase of $1\%$ is needed.  The rates for CFHT observations of
M31 are shown in figure \ref{fig:m31}.

In figures \ref{fig:peak}--\ref{fig:starstar} and \ref{fig:m31}, we plot
$d\Gamma/d\ln\tf$.  This is the differential event rate for events of a
specific $\tf$.  To get the total event rate, we must integrate this over the
range of $\ln\tf$ that the sampling strategy is sensitive to.  Roughly
speaking, we can observe events with $\tf$ between about the twice the sample
spacing and the total monitoring duration.

\section{Feasibility of Einstein Time Measurement}

We now discuss the possibility of using the statistical timescale method to
determine the Einstein time.  We know that we will need a high signal to noise
for any method to work.  We specialize to the case of a pixel lensing
experiment that may resolve a great number of stars, perhaps even enough for a
classical microlensing survey.  This is the case, for example, when using the
new 12288x8192 pixel CCD on the CFHT to observe M31.  The pixel scale is
$0.\arcsec 2$, giving a substantial field of view, with a large number of
resolved stars in a modest integration time.  With this type of data, the rate
of high signal to noise events on unresolved stars is quite high, thus we might
hope to be able to recover the Einstein times of a subset of the pixel events.

Measuring even the truncated statistical timescale $\tst$ is quite difficult.
We now investigate the real-world possibility of learning anything from this
measure.  We perform Monte-Carlo simulations of the observed $\tst$ versus the
true $\tst$ to determine the errors involved in this process.

We proceed as follows.  For a given lightcurve in a dataset, we can easily
measure $\tf$, which also tells us to what value $n\tf$ the lightcurve is well
measured.  We can then fit a constant background to the lightcurve and
determine $\tst$.  We have the theoretical curve for $\tst$, which lets us read
off the minimum impact parameter $\beta$, which gives us the full information
on the microlensing event.

We would like to know what the error in the determination of $\beta$ is.  From
the lightcurve, we can determine the signal to noise in the peak.  We can use a
Monte-Carlo simulation to generate events at that fixed signal to noise, for
different values of beta.  For each generated lightcurve, we compute $\tst$,
and thus we generate a histogram of measured values, which allows us to
determine the error in the measured $\tst$, thus an error in the measured
$\beta$.

As a concrete case, we take microlensing lightcurves with peak signal to noise
$Q_0=42$, six times the basic detection threshold.  We fix $\tf=3$ days, take
exposures of 3600 seconds, and vary combinations of the minimum impact
parameter $\beta$, and the apparent magnitude $m_V$ that give our chosen peak
signal to noise.  Observation are taken once per day for thirty days on either
side of the maximum.  In figure \ref{fig:lc}, we show generated lightcurves for
these types of events.  For $\delta=20$ and $\delta=100$, we give example
lightcurves with $\tf=3$ days, for peak signal to noise of $Q_0=7$ and
$Q_0=42$.  The values of $\beta$ that we choose correspond to
$\delta=(10,20,50,100)$ in the peak.  The corresponding absolute $V$-magnitudes
are $M_V=(1.78,2.53,3.42,4.28)$.  We now perform a Monte-Carlo simulation in
which we generate events of these fixed parameters and measure directly the
ratio $\xi_{10}=t_{\sigma 10}/\tf$.  With these parameters, we find that the
four values of $\beta$ are distinguishable at the $1\sigma$ level.  In figure
\ref{fig:hist}, we plot the event histograms for these cases.

For this technique to work, it is important that a good determination of the
background is made.  We will assume that the background is determined to
$1\sigma$ by fitting a degenerate lightcurve.  The error will be approximately
the square root of the number of counts, times the factor of two to account for
non-poissonian errors, and divided by the square root of the number of samples.
We find that this error in the determination of the background induces an error
in the determination of $\xi_{10}$ of $\pm0.029$, considerably smaller than the
other errors.  Thus, the error in the background determination is negligible.

\section{Discussion and Conclusions}

The flux-weighted full-width at half maximum timescale can be used in pixel
microlensing experiments to measure the optical depth.  This determination does
not require the almost degenerate fit to the lightcurve, as was necessary with
previous methods.  The optical depth is a strong indicator of the amount of
mass in lenses along a line of sight.  With this knowledge, we can further
investigate the properties of the lenses.

It is clear from figures \ref{fig:peak}--\ref{fig:starstar},\ref{fig:m31} that
the shape of the timescale distribution of events depends strongly on the mass
of the lenses.  If a suitable population of events can be detected, it will be
possible to make a rough estimate of the mass of the intervening lenses.  This
will serve to provide hints as to the nature of the lenses, which is unknown.
For example, a large population of solar--mass lenses, inferred to be neutron
stars, may be difficult to reconcile with the metallicity of the galactic gas.
Such a population might be assumed to be primordial black holes.  A
sufficiently large optical depth of any mass lens may also be difficult to
reconcile with the total baryon density of the universe.  In contrast,
microlensing is a probe of the low--mass end of the stellar luminosity
function, and may give some indication of the number and mass of brown dwarfs
in galactic systems.

With a measurement of the optical depth in hand, measuring the Einstein times
of a subset of the lightcurves provides a great deal more information.  This is
a more discriminatory determination of the masses of the lensing population.
Even if only a small subset of events can be so analyzed, the added information
is very instructive.

In conclusion, we have expanded the theory of pixel microlensing in light of
the result of Gondolo (1999) that a flux--weighted timescale can be used to
unambiguously determine the optical depth even with moderate quality data.  We
have described how to compute event rates in terms of the new timescales, and
applied the techniques to several proposed microlensing surveys.  Lastly, we
have described a new method by which the Einstein time can be determined from
high quality data.  This method is complementary to Gould's (1996) method.
Though complicated to implement, it is perhaps conceptually simpler.  The
physical parameters are determined not through a lightcurve fit, but through a
comparison of the widths of the peak and the tails.

\acknowledgements

E.B. thanks Paolo Gondolo, Gordon Squires and Tom Broadhurst for useful
conversations.  We thank Yannick Giraud-H\'{e}raud, Jean Kaplan, Tod Lauer,
Harvey Richer, Mike Shara, and Steve Zepf for suggestions.  This research was
supported by grants from NASA and DOE.

\newpage

\begin{figure}
\caption{The dependence of the peak impact parameter on the ratio
$\tst/\tf=\xi_n/w$ for several values of $n$.  This ratio varies significantly
for $\beta$ between a few tenths and a hundredth.  Thus it may be feasible to
make a crude measurement of $\beta$ from measurements of $\tst$ and $\tf$.}
\label{fig:tsigma}
\end{figure}
\noindent
\epsfig{width=6.5in,file=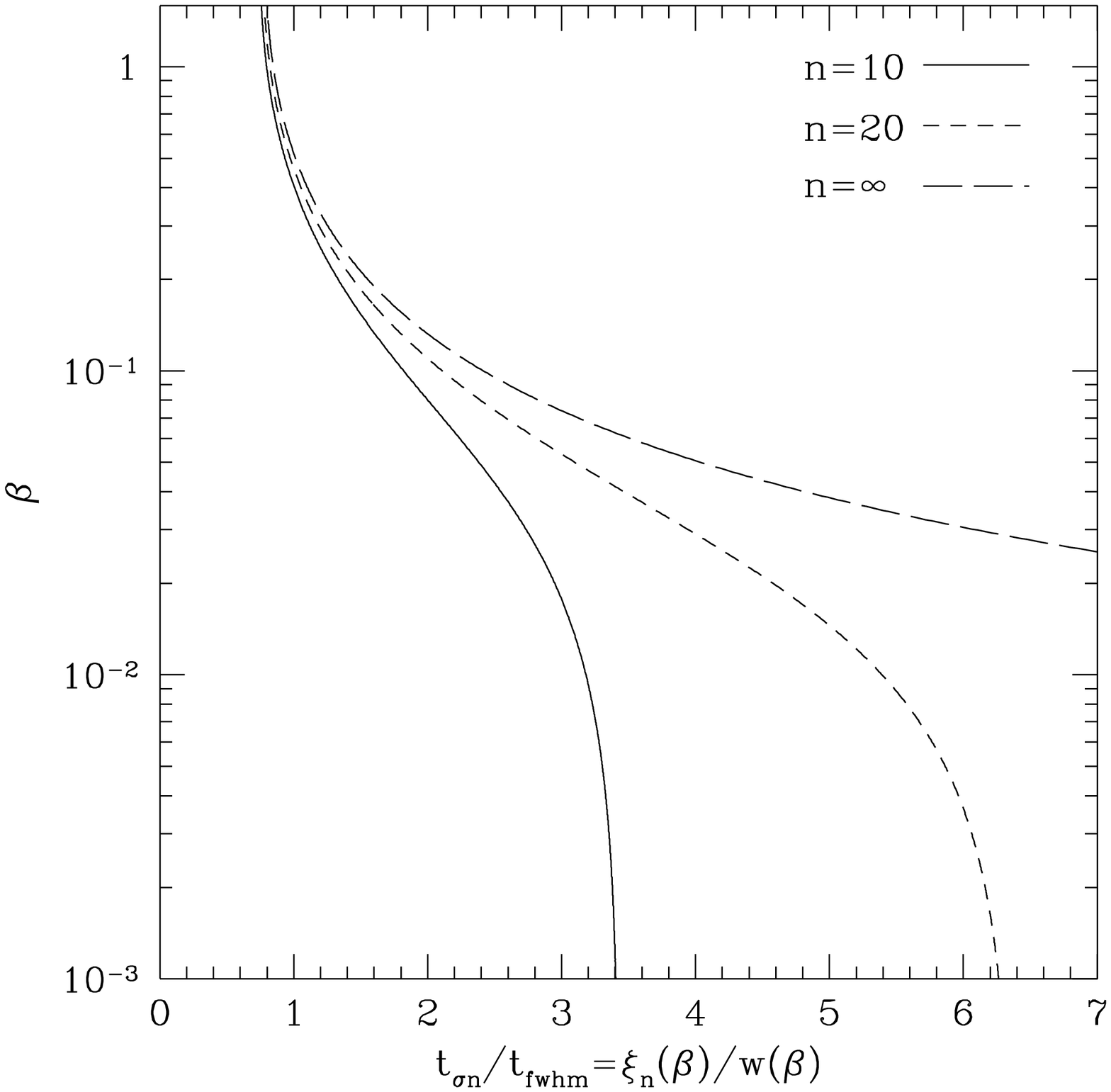}

\newpage
\begin{figure}
\caption{Pixel lensing rates for the peak threshold trigger on WFPC2, AC, and
NGST.  We have taken $Q_0=7$.  In the bottom right we plot rates in the
Advanced Camera with $Q_0=42$.}
\label{fig:peak}
\end{figure}
\noindent
\epsfig{width=3.2in,file=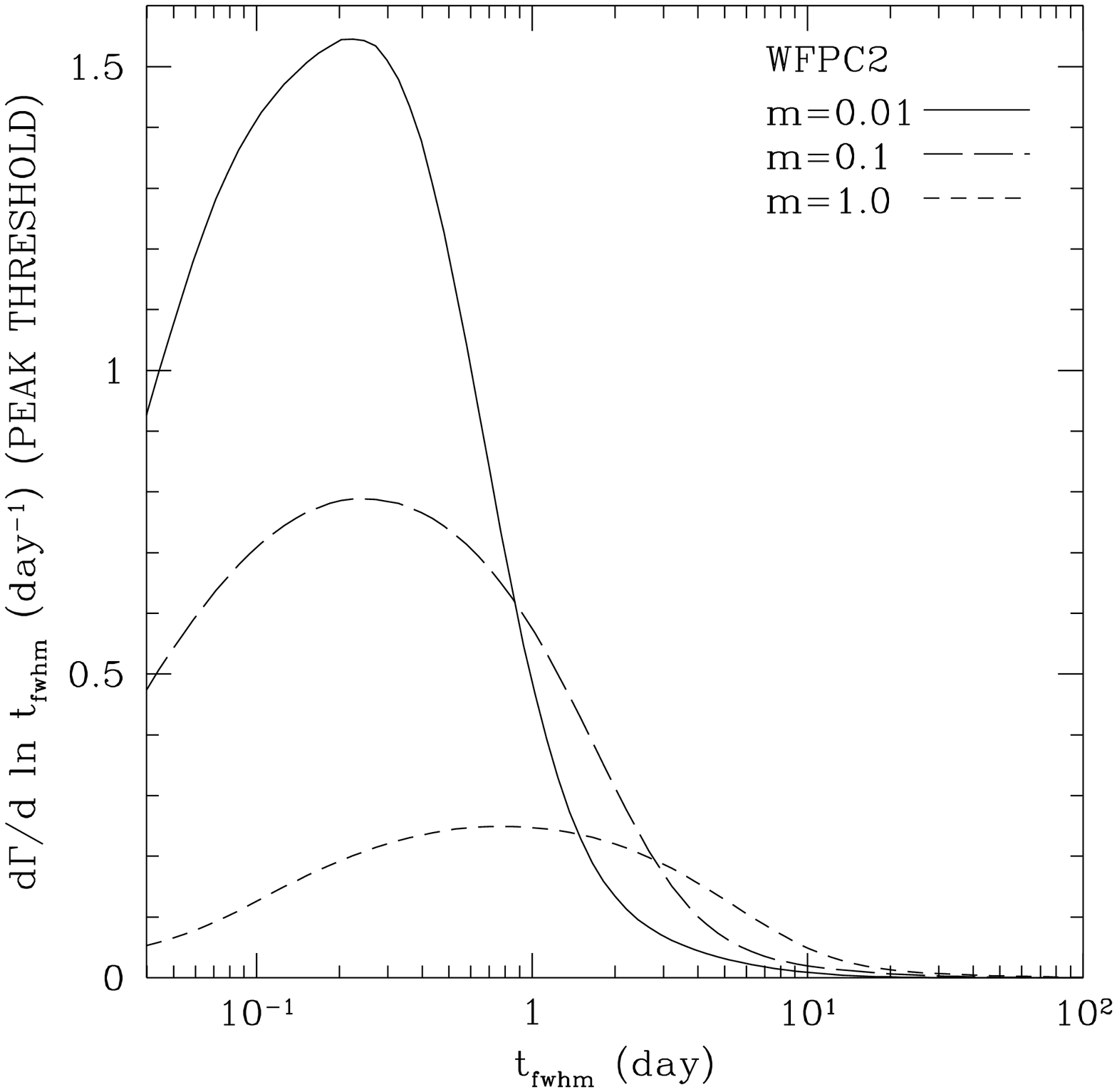}
\epsfig{width=3.2in,file=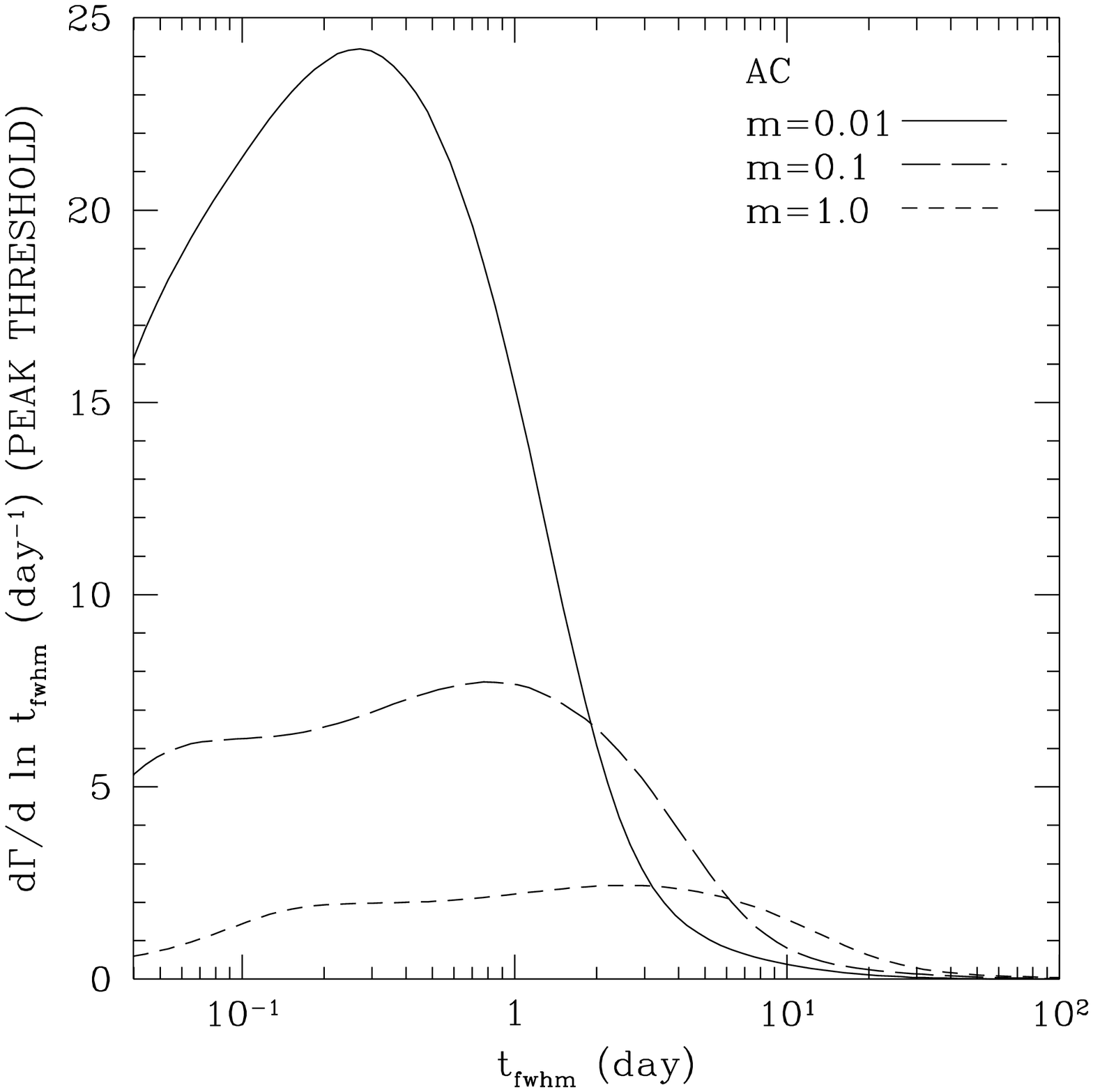}\\
\epsfig{width=3.2in,file=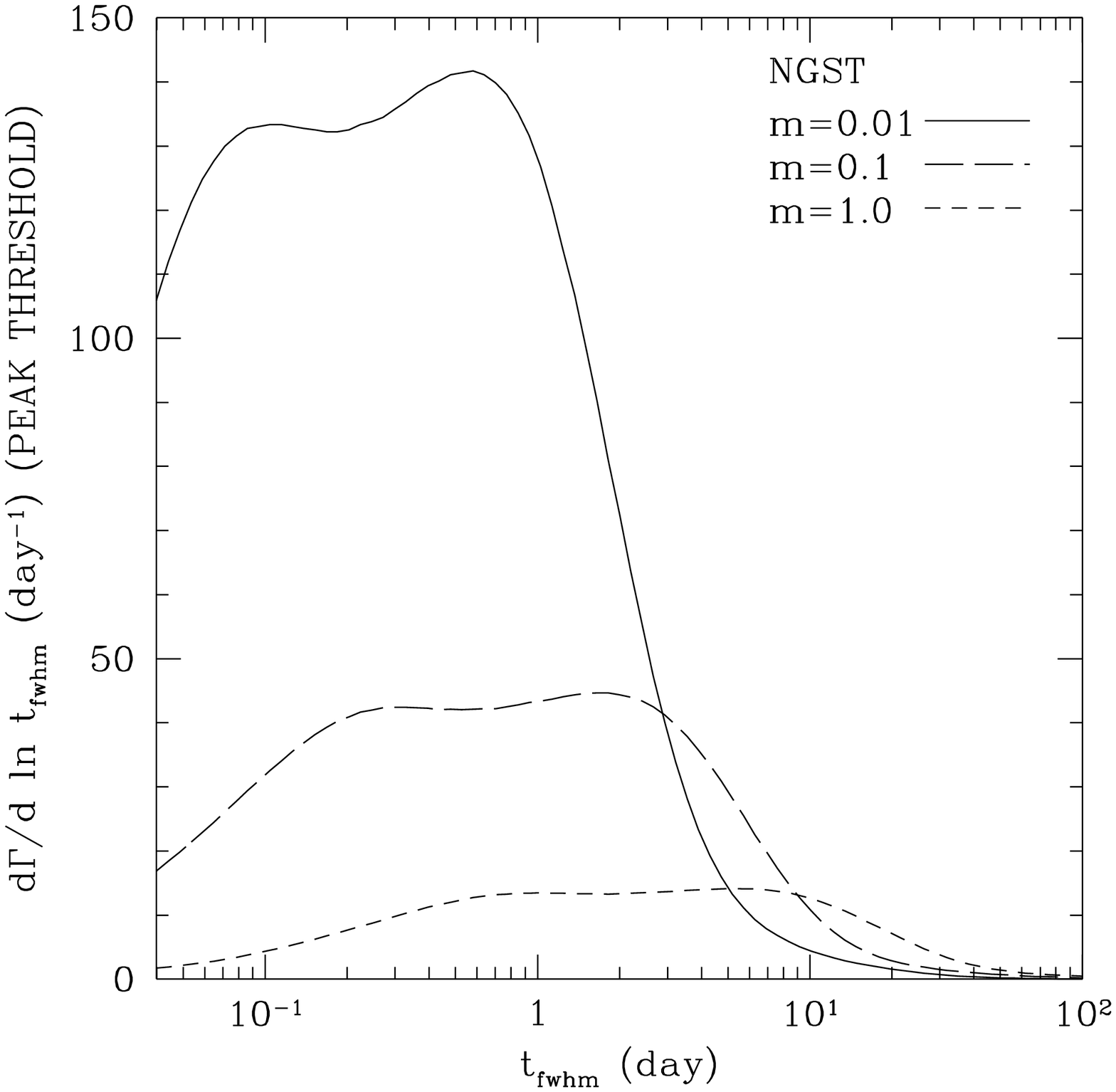}
\epsfig{width=3.2in,file=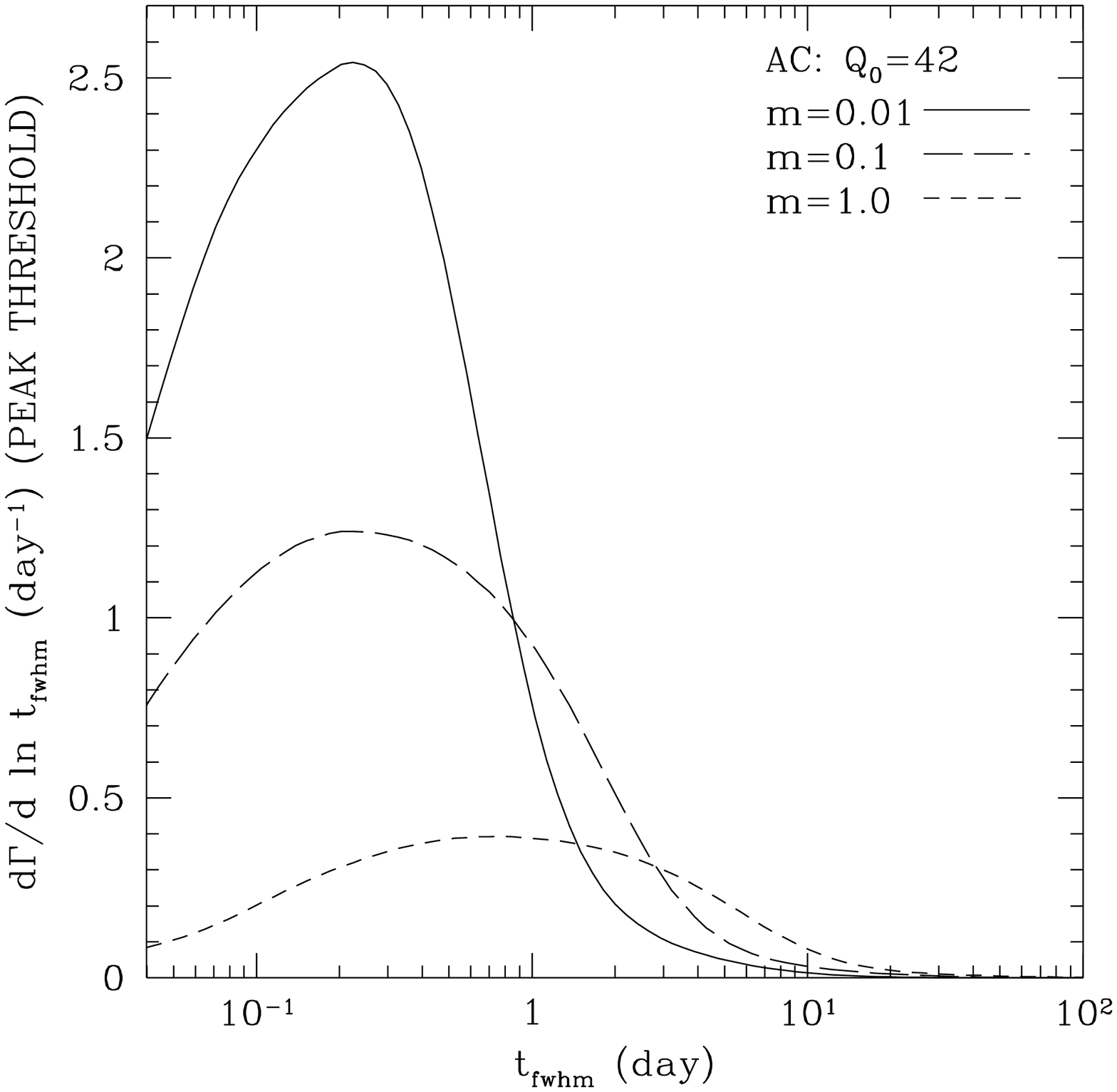}

\newpage
\begin{figure}
\caption{Pixel lensing rates for the improved peak threshold trigger on WFPC2,
AC, and NGST.  The spacing of observations is one day, except in the bottom
right plot, where it is six hours.}
\label{fig:peak3}
\end{figure}
\noindent
\epsfig{width=3.2in,file=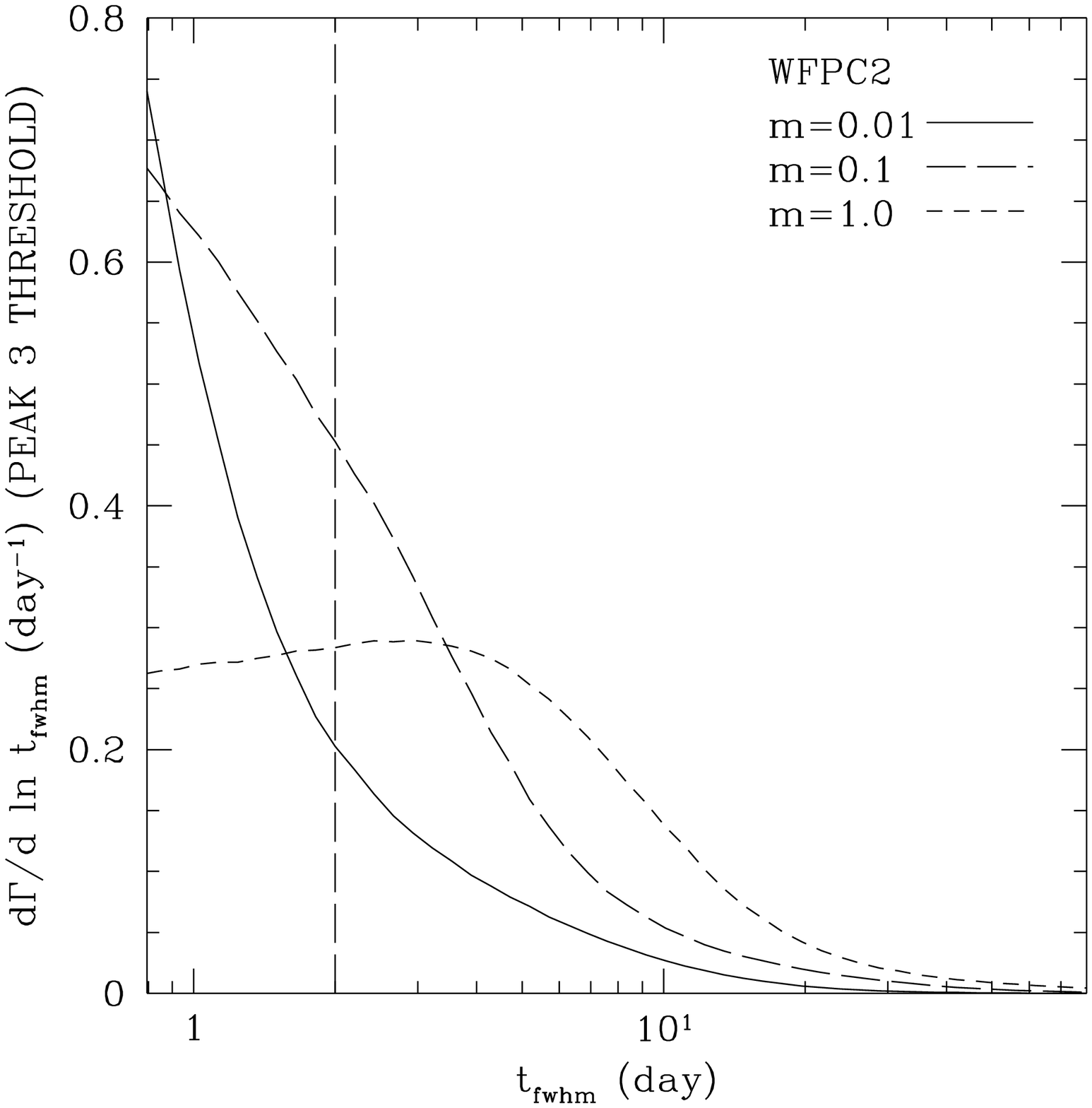}
\epsfig{width=3.2in,file=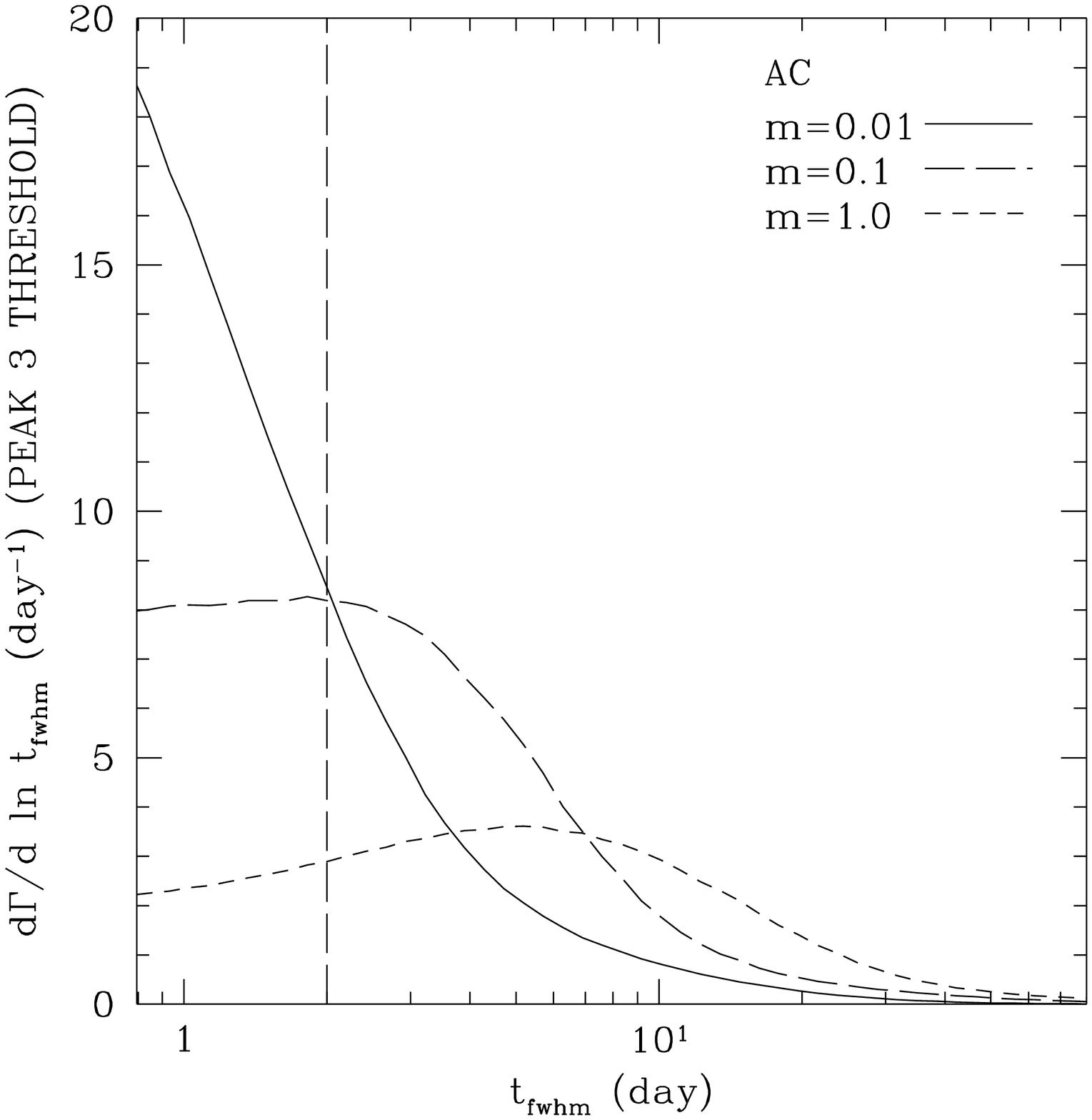}\\
\epsfig{width=3.2in,file=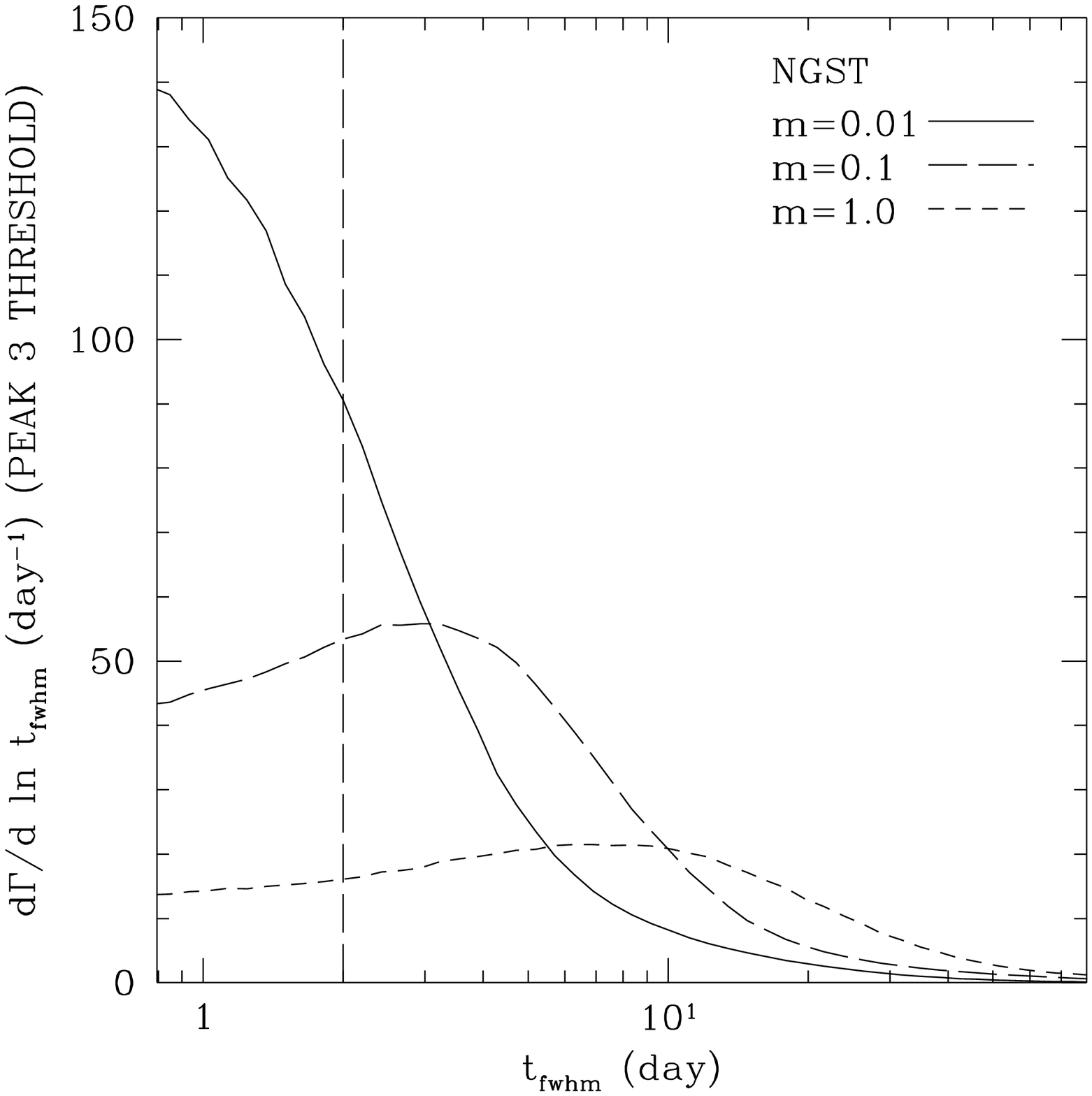}
\epsfig{width=3.2in,file=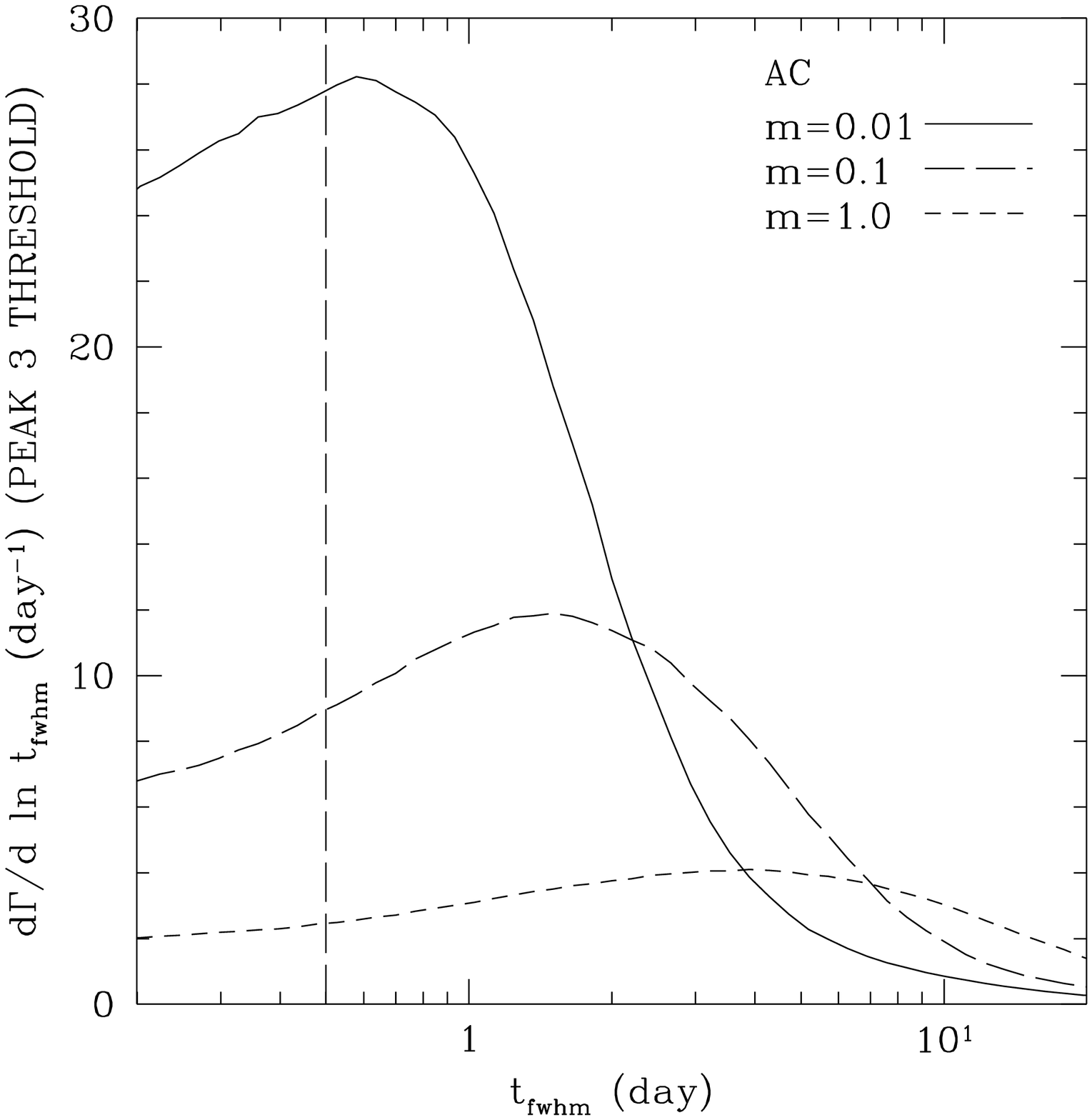}

\newpage
\begin{figure}
\caption{Pixel lensing rates for the event threshold trigger on WFPC2, AC, and
NGST.  The spacing of observations is one day, except in the bottom right plot,
where it is six hours.}
\label{fig:event}
\end{figure}
\noindent
\epsfig{width=3.2in,file=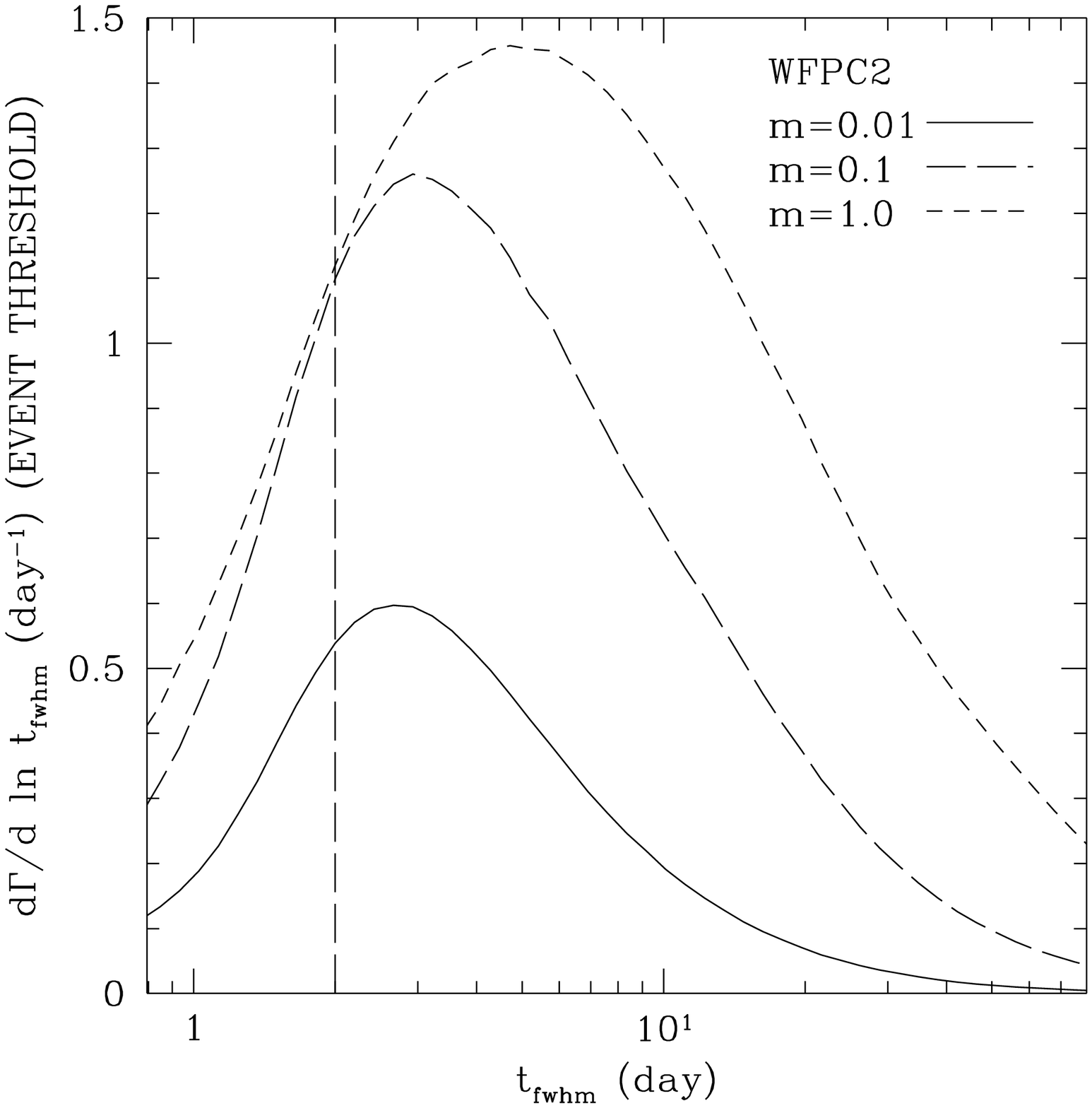}
\epsfig{width=3.2in,file=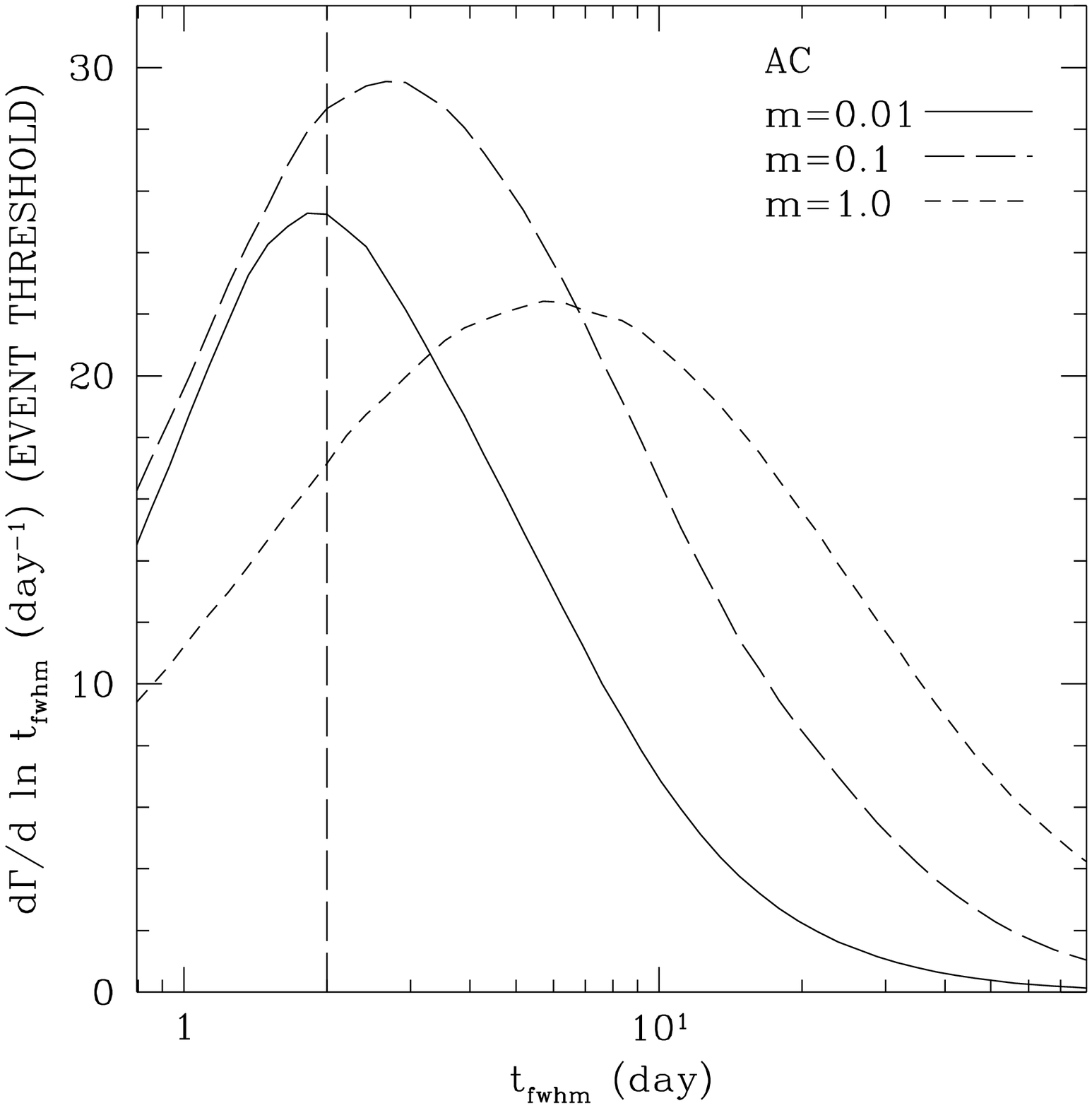}\\
\epsfig{width=3.2in,file=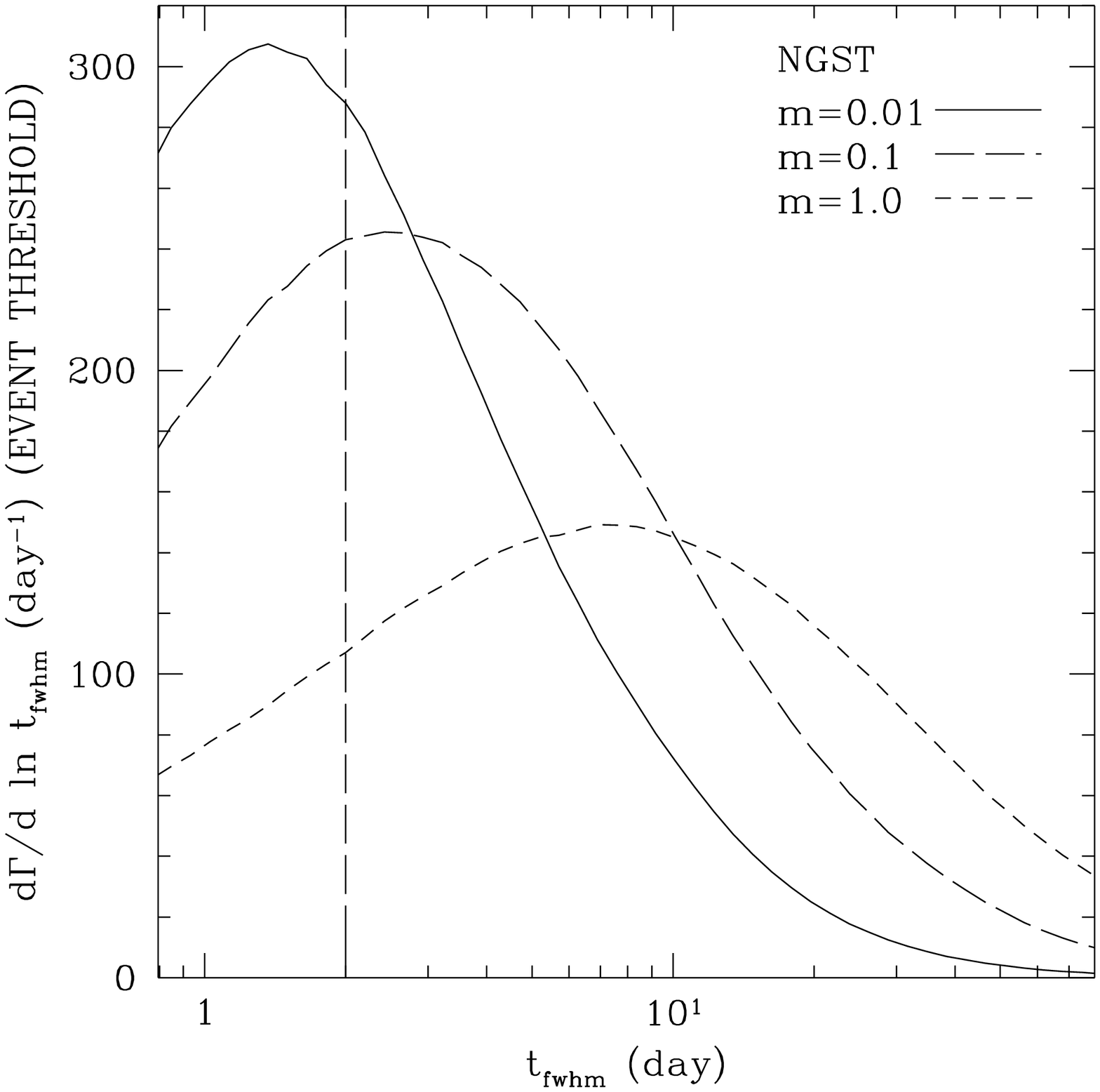}
\epsfig{width=3.2in,file=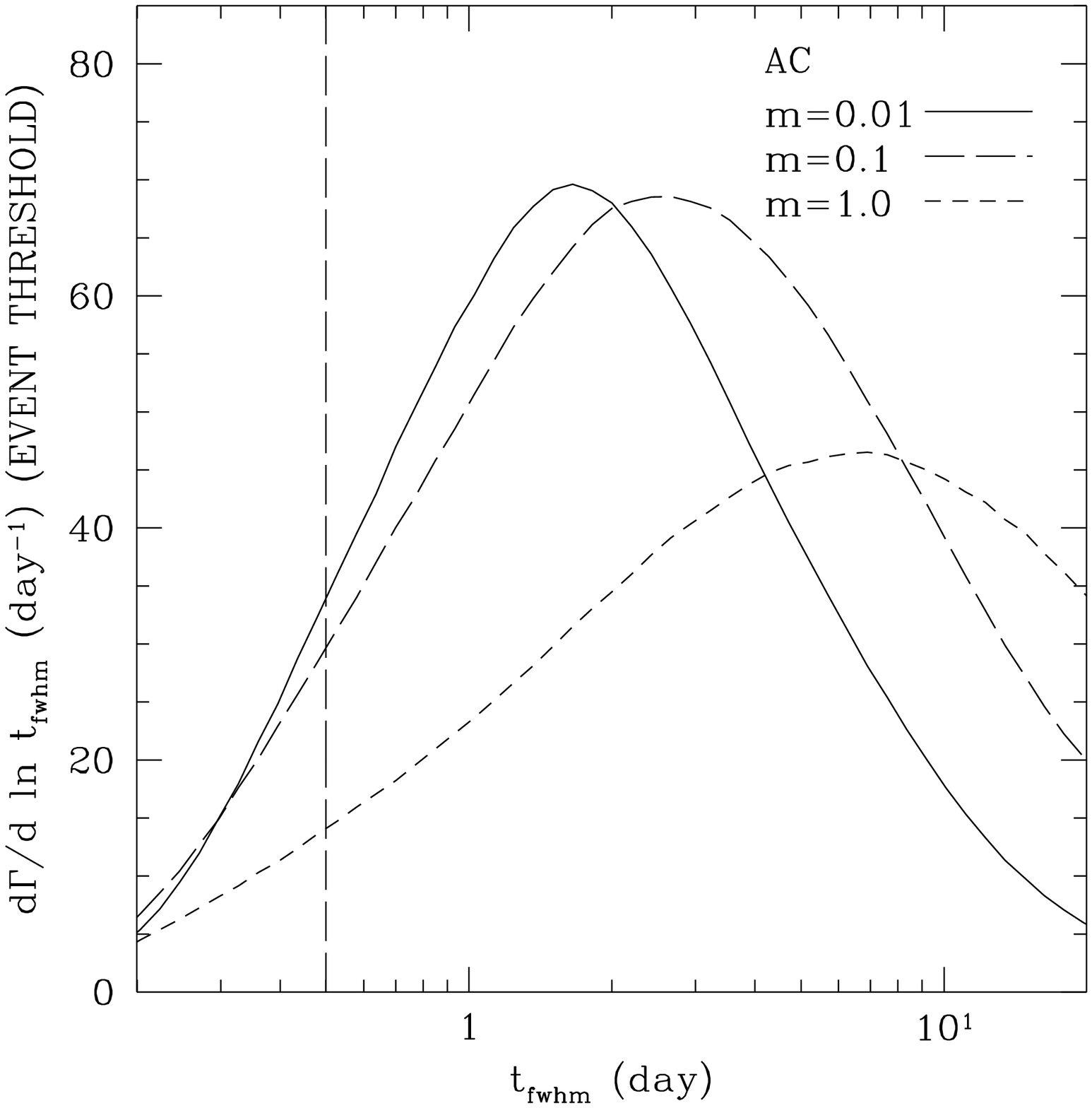}

\newpage
\begin{figure}
\caption{Star--star lensing rates for the improved peak threshold trigger on
WFPC2, AC, and NGST.  The spacing of observations is one day, except in the
bottom right plot, where it is six hours.  We have plotted both a typical
estimate of the average star mass, 0.4 $M_\odot$, along with an extreme
estimate, implying a great number of brown dwarfs, 0.05 $M_\odot$.}
\label{fig:starstar}
\end{figure}
\noindent
\epsfig{width=3.2in,file=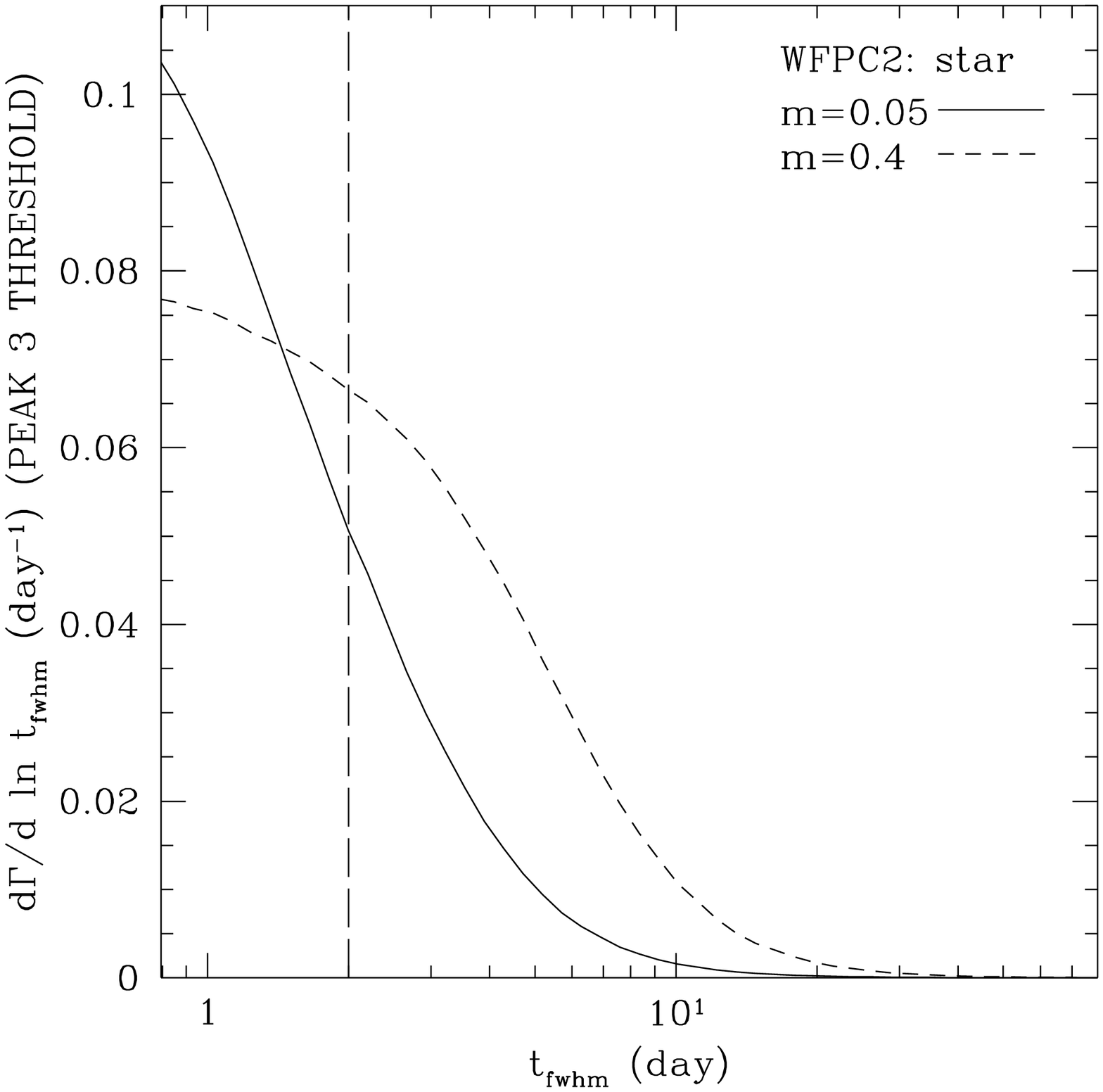}
\epsfig{width=3.2in,file=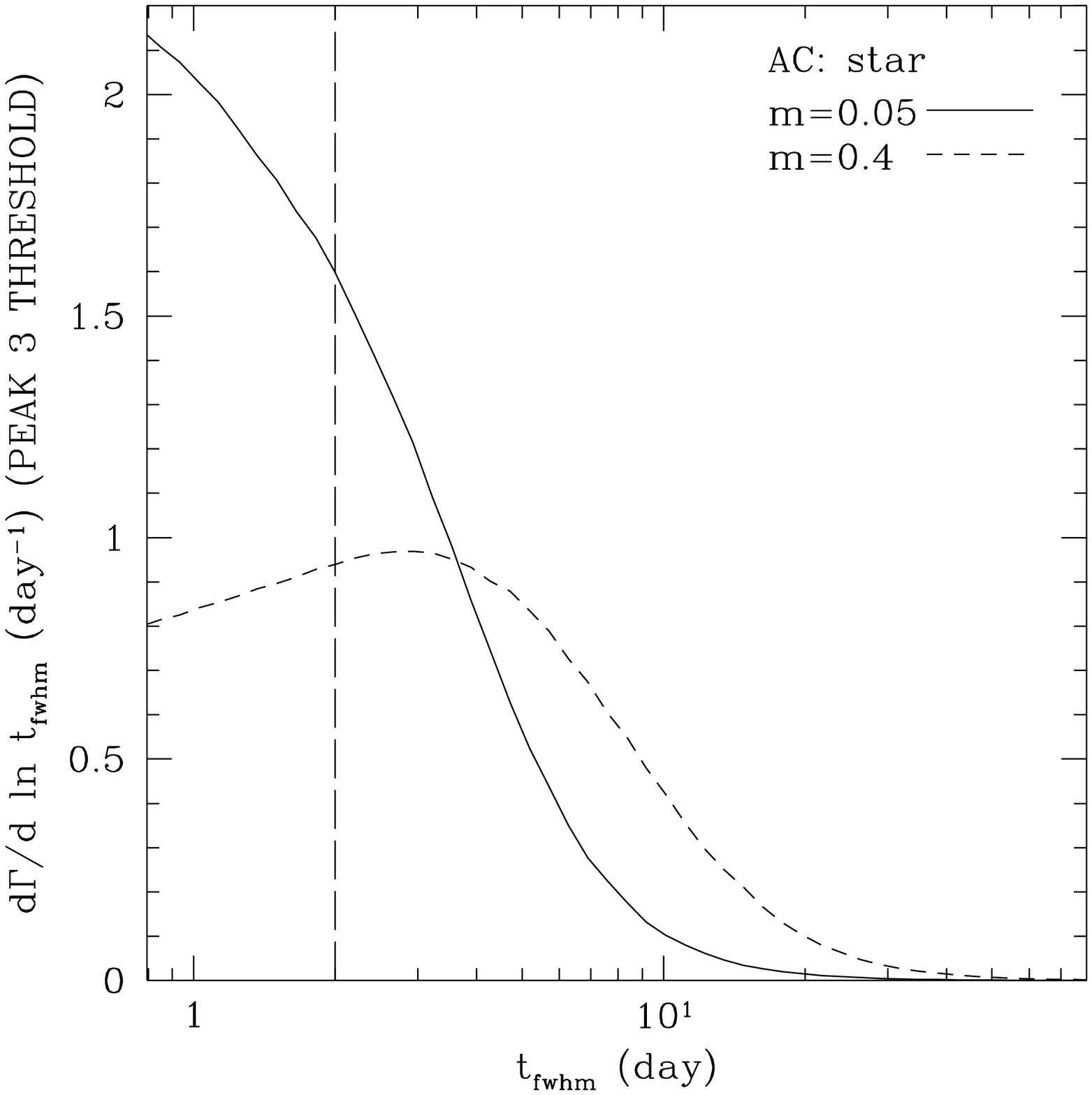}\\
\epsfig{width=3.2in,file=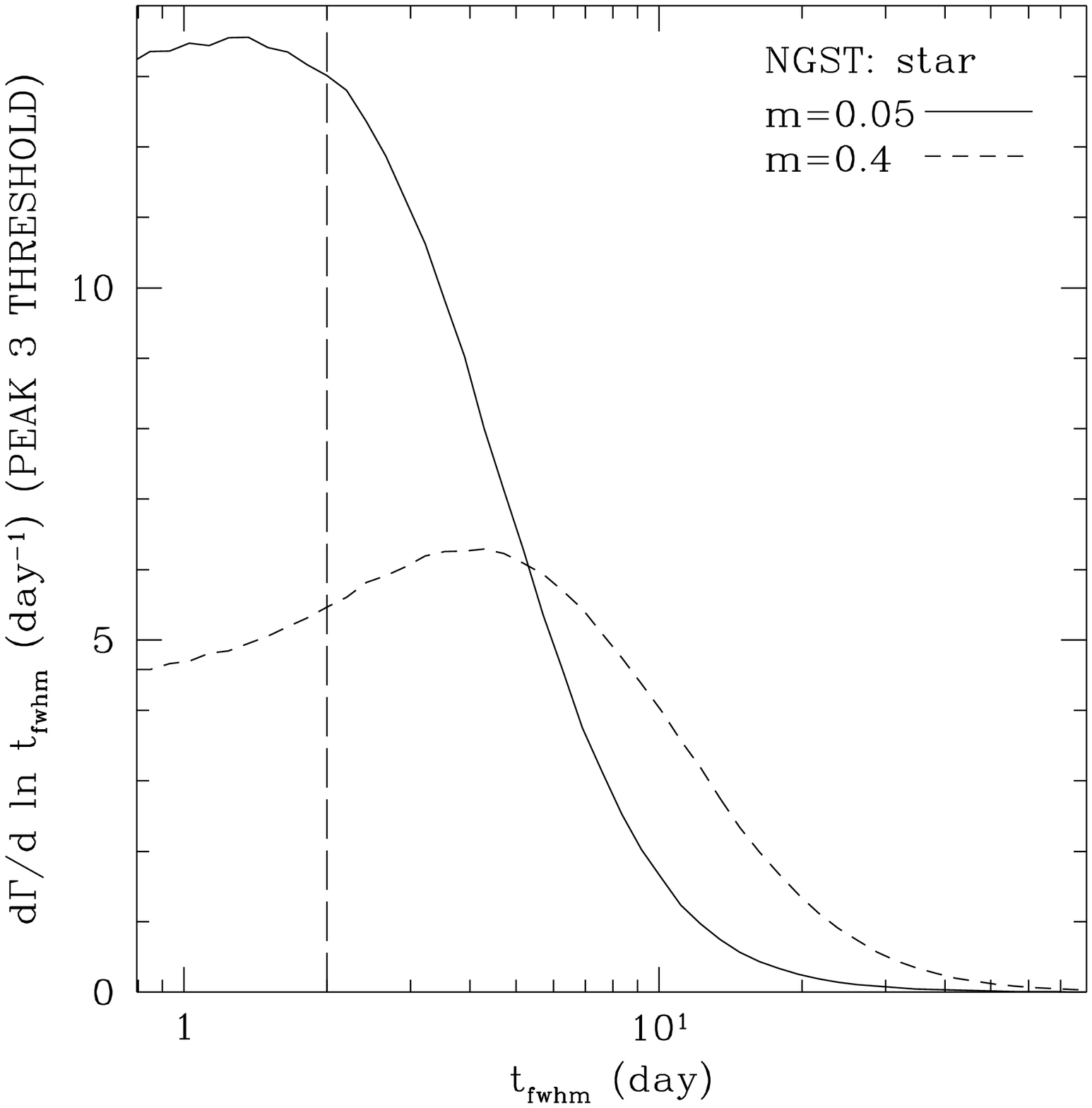}
\epsfig{width=3.2in,file=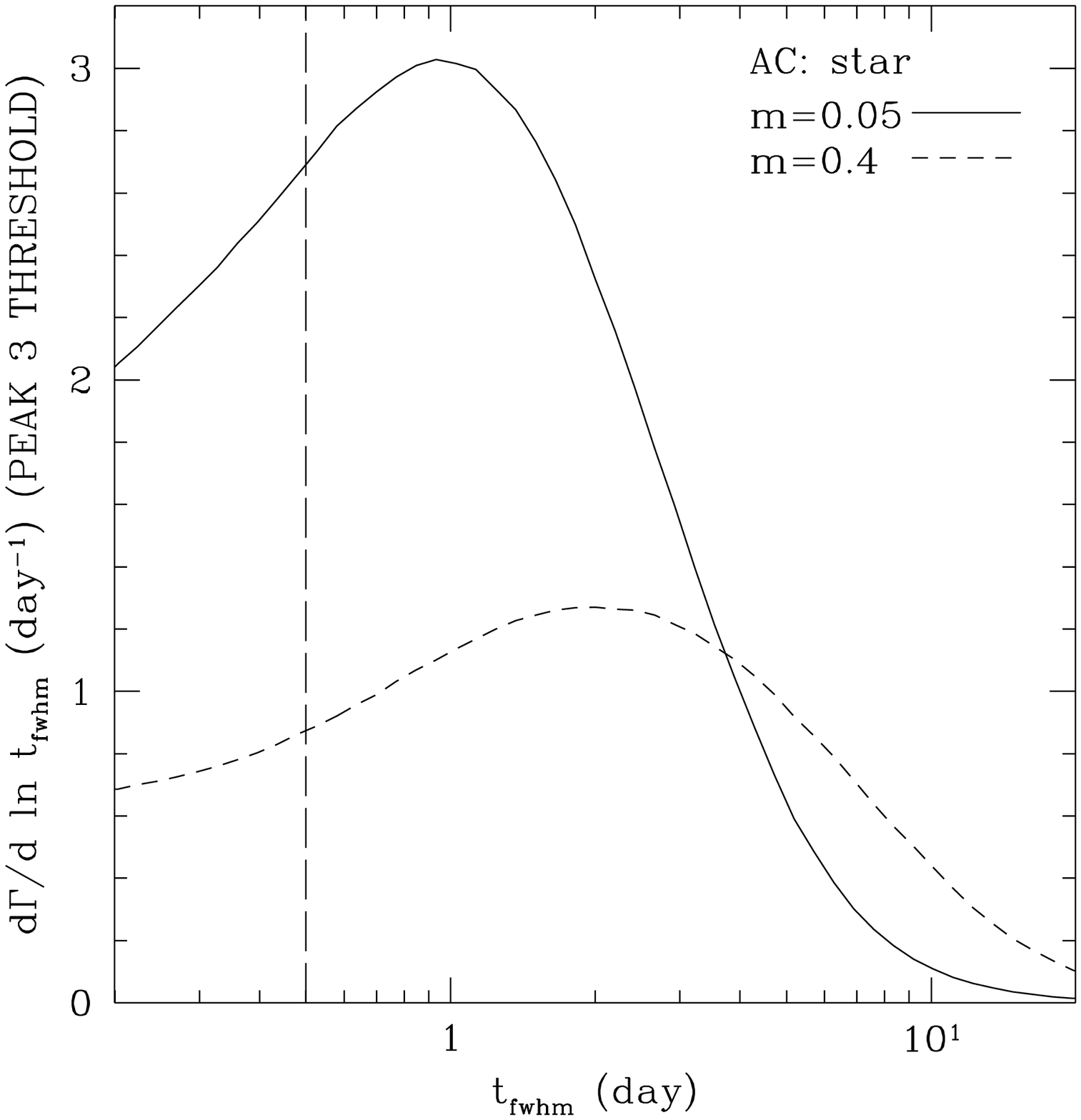}

\newpage
\begin{figure}
\caption{Contours of constant optical depth to the disk of M31.  These are
given in (dimensionless) units of $10^{-6}$.  The isophotes (thick lines) are
taken from Hodge \& Kennicutt (1983), and are 19, 20, 21, 22, 23, 24, 25 $B$
magnitudes per square arc second.  The rectangular boxes show the field of view
of the CFHT12k CCD, centered on M31 and aligned along both the major and minor
axes.}
\label{fig:contour}
\end{figure}
\noindent
\epsfig{width=7in,file=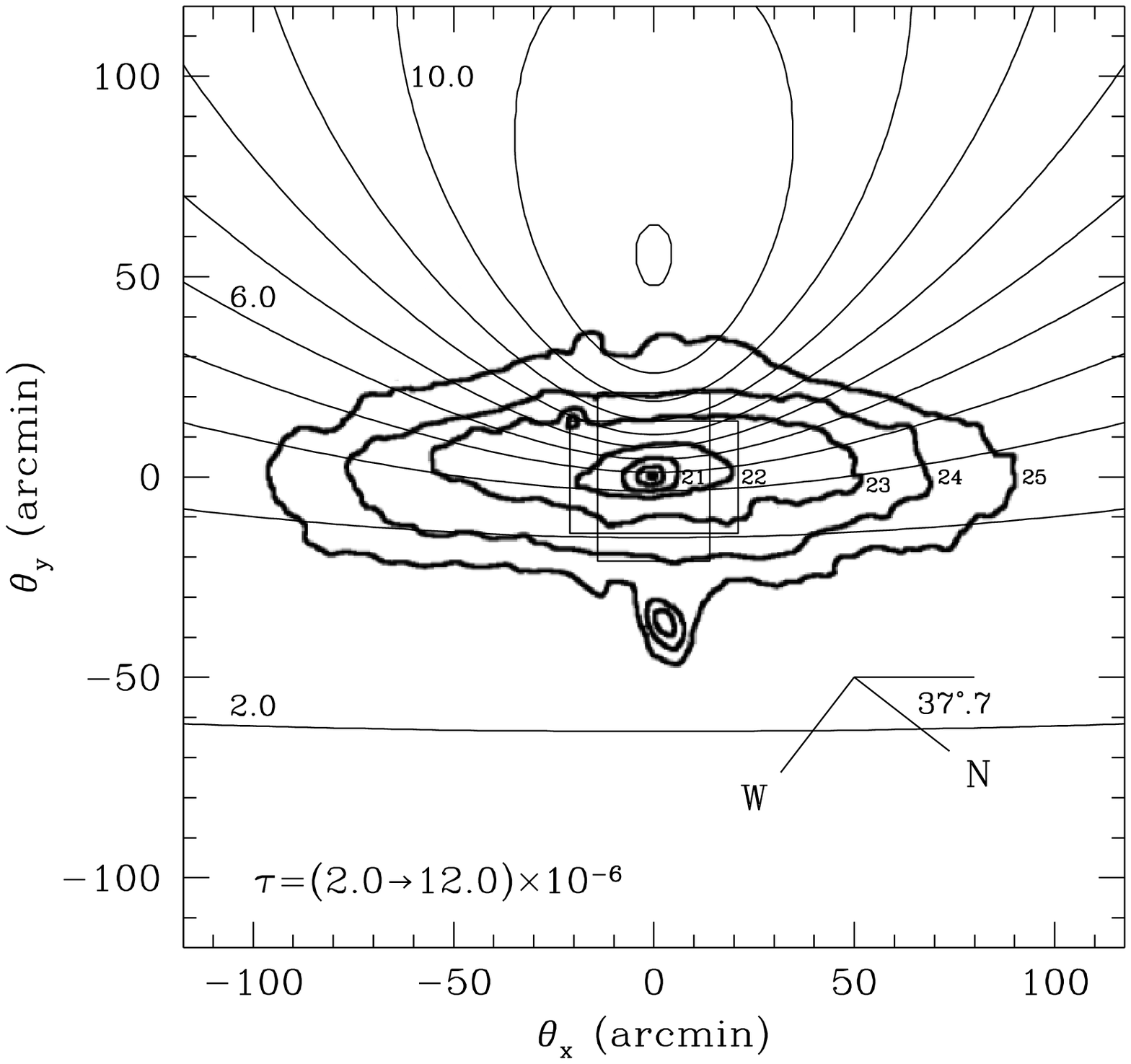}

\newpage
\begin{figure}
\caption{Pixel lensing rates for M31 using the CFHT 12k$\times$ 8k CCD.  The
top left shows the peak threshold trigger with $Q_0=7$.  The top right shows
the improved peak threshold trigger, also with $Q_0=7$.  The bottom left shows
the improved peak threshold trigger with $Q_0=42$, six times larger than the
other cases.  The bottom right shows the rate of events where the source star
is resolved.  The spacing of observations is one day.}
\label{fig:m31}
\end{figure}
\noindent
\epsfig{width=3.2in,file=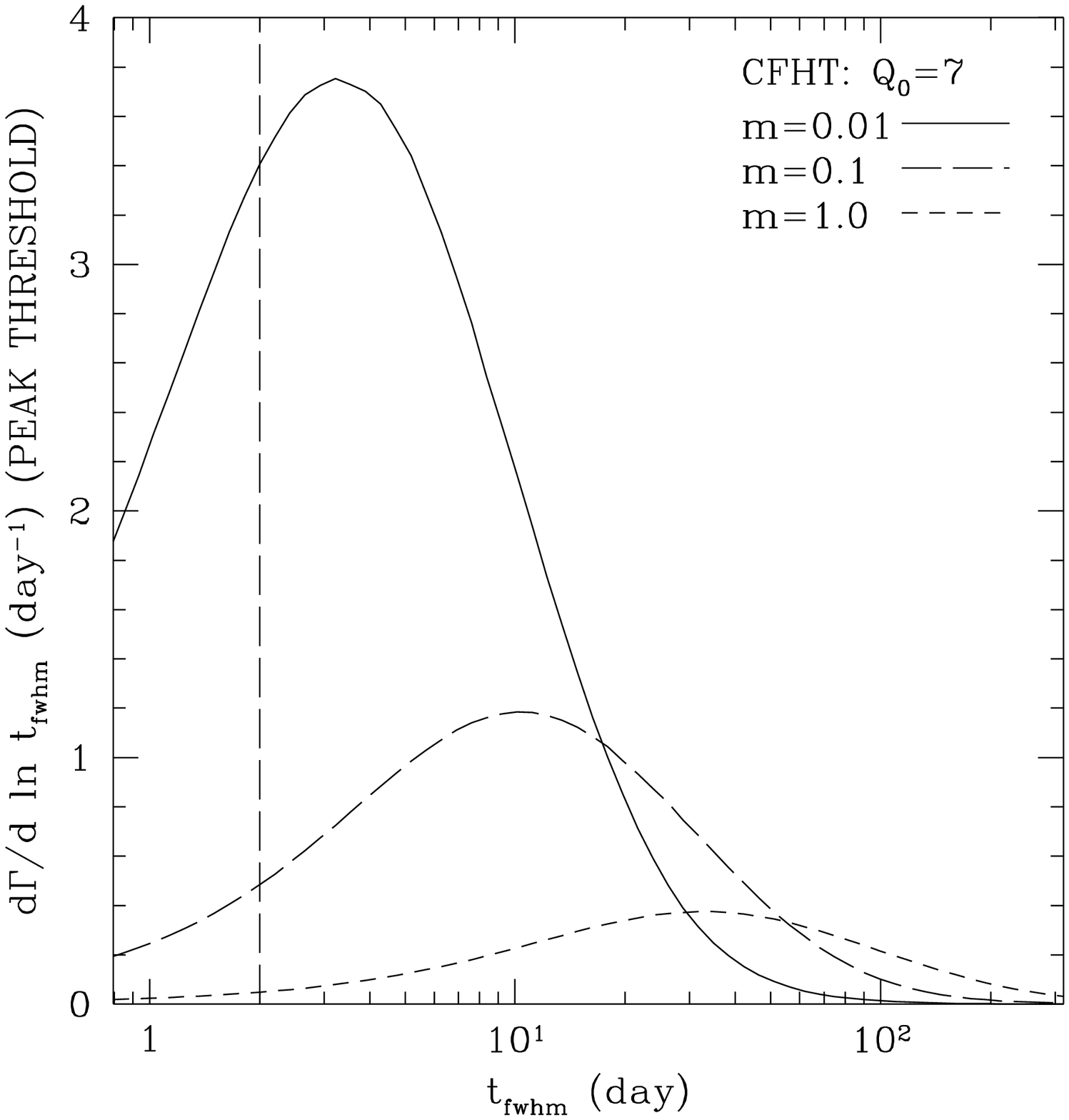}
\epsfig{width=3.2in,file=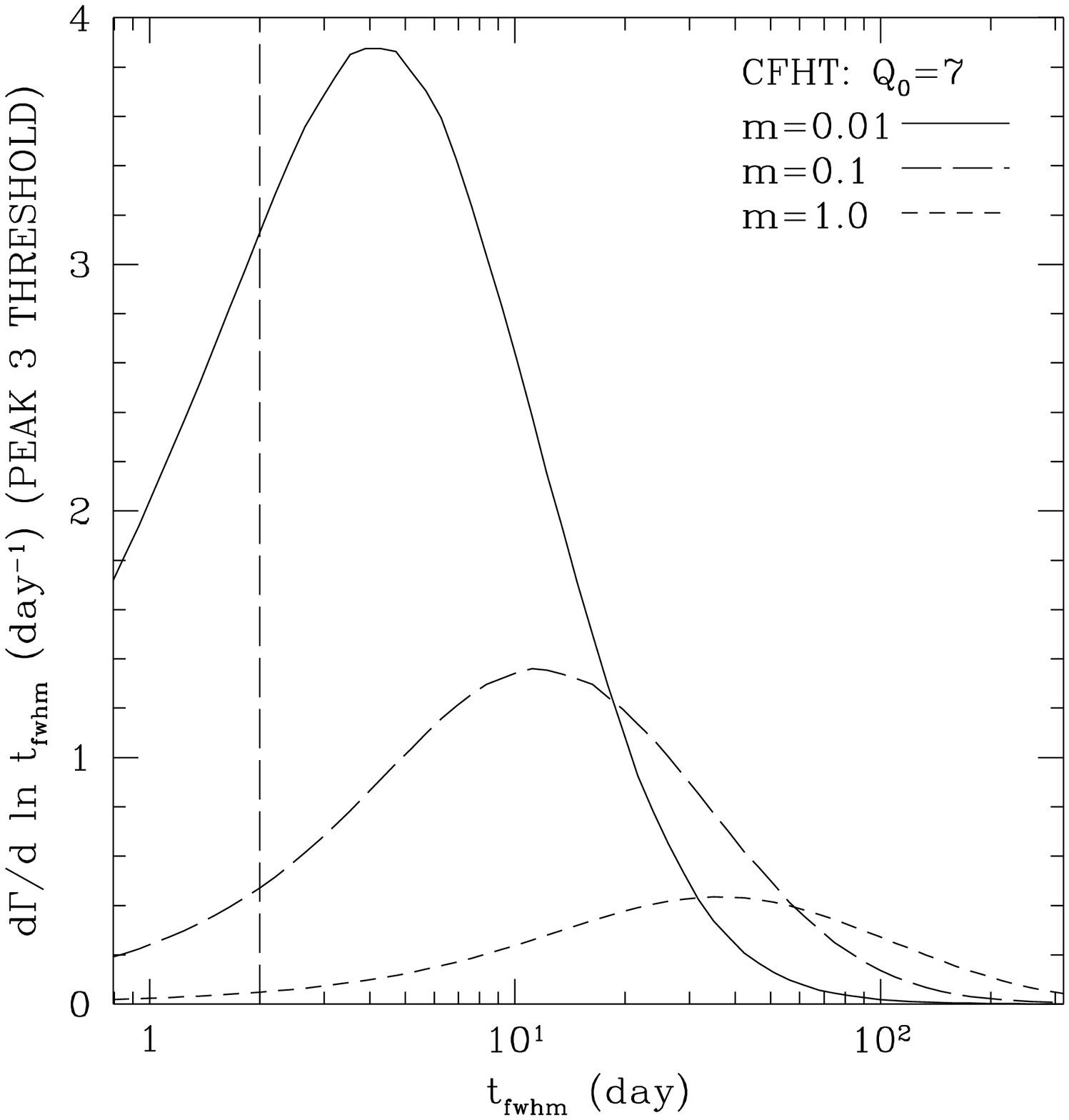}\\
\epsfig{width=3.2in,file=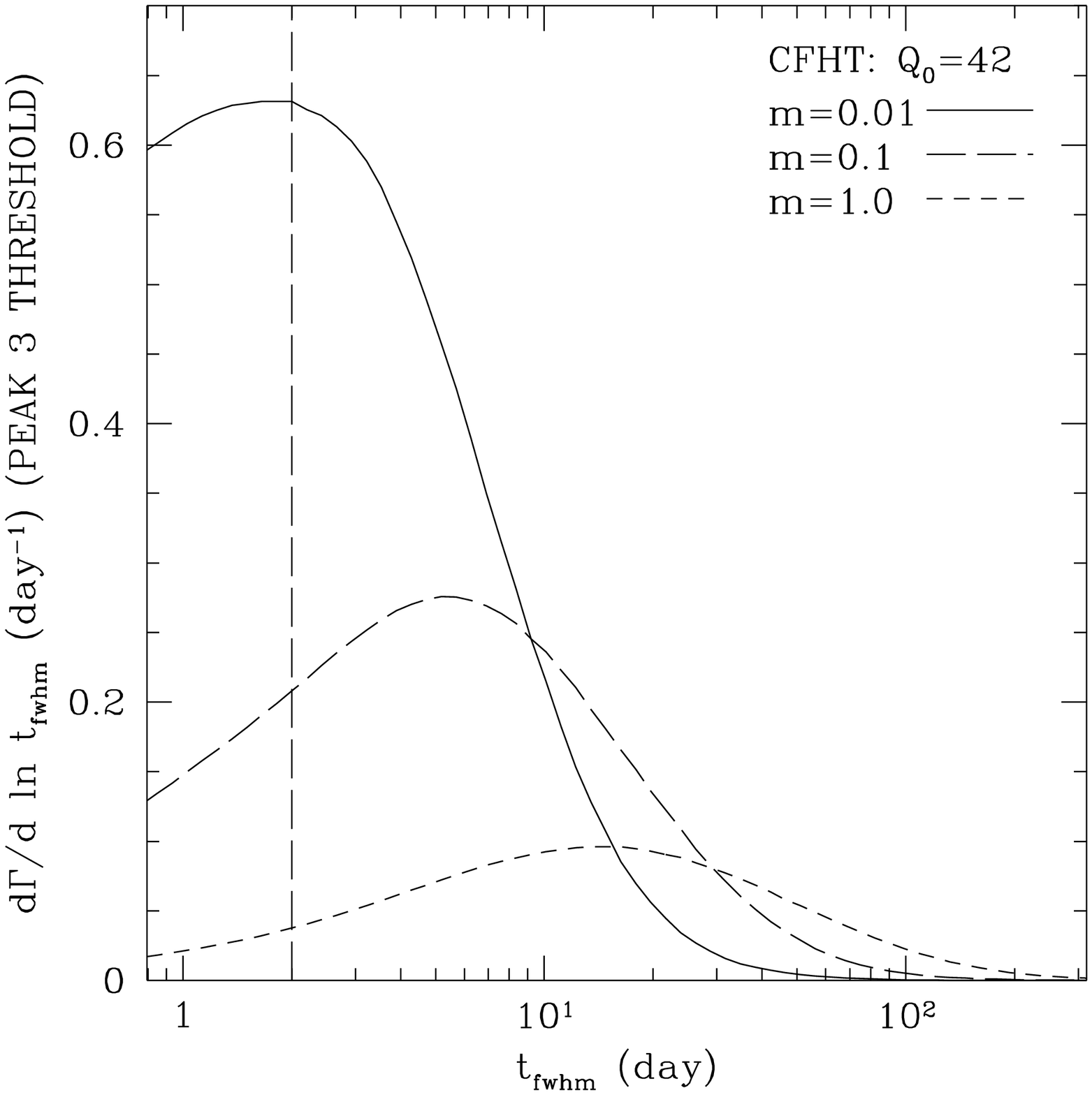}
\epsfig{width=3.2in,file=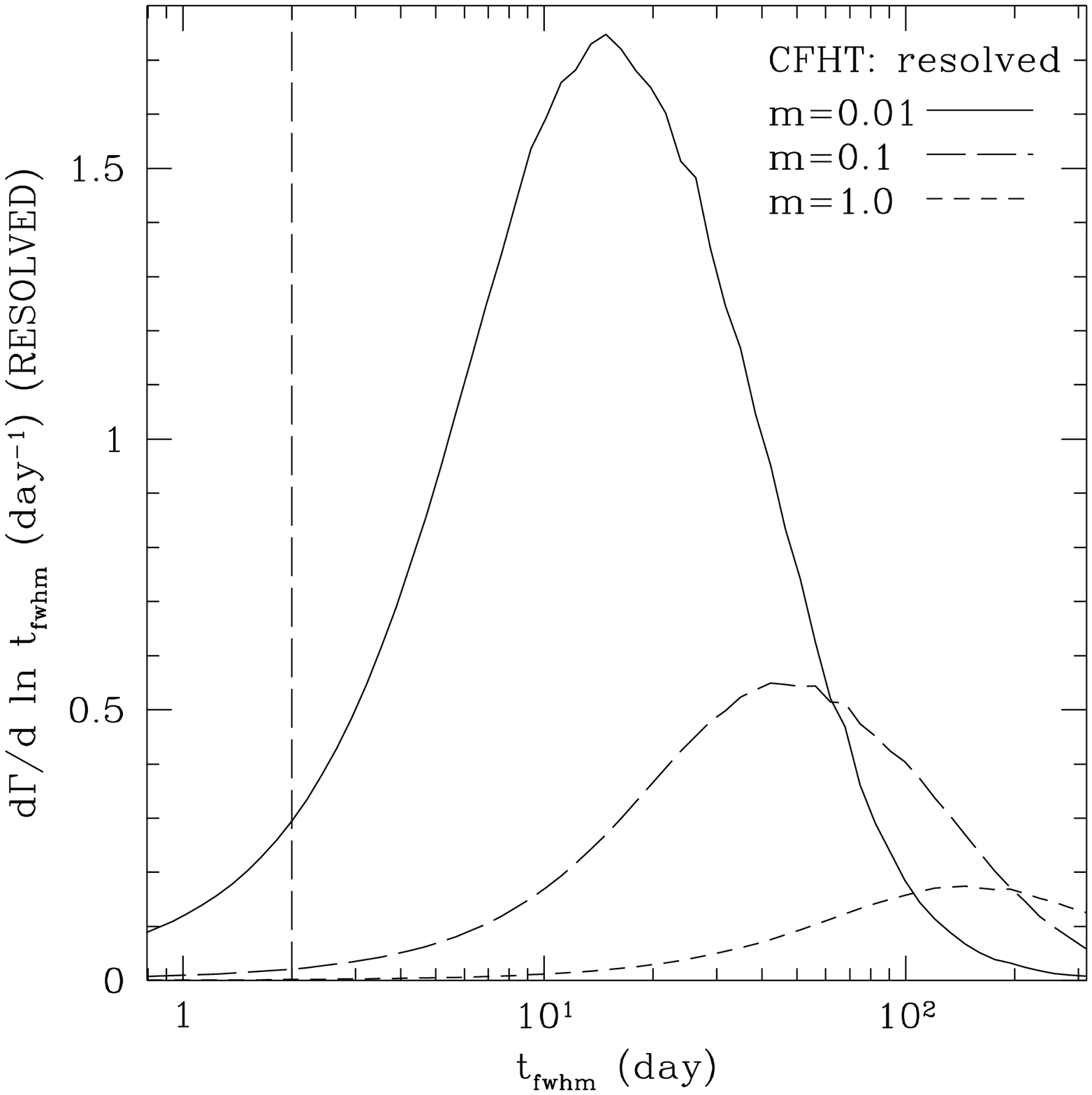}

\newpage
\begin{figure}
\caption{Sample lightcurves for M31 events in CFHT.  The dashed lines indicate
the background level and the prescribed detection levels, $7\sigma$ and
$42\sigma$, respectively.  The solid lines are the true lightcurves, and are
not fits.
{\em Top left}: $Q_0=7$,
$\delta=20$.
{\em Top right}: $Q_0=7$,
$\delta=100$.
{\em Bottom left}: $Q_0=42$,
$\delta=20$.
{\em Bottom right}: $Q_0=42$,
$\delta=100$.}
\label{fig:lc}
\end{figure}
\noindent
\epsfig{width=3.2in,file=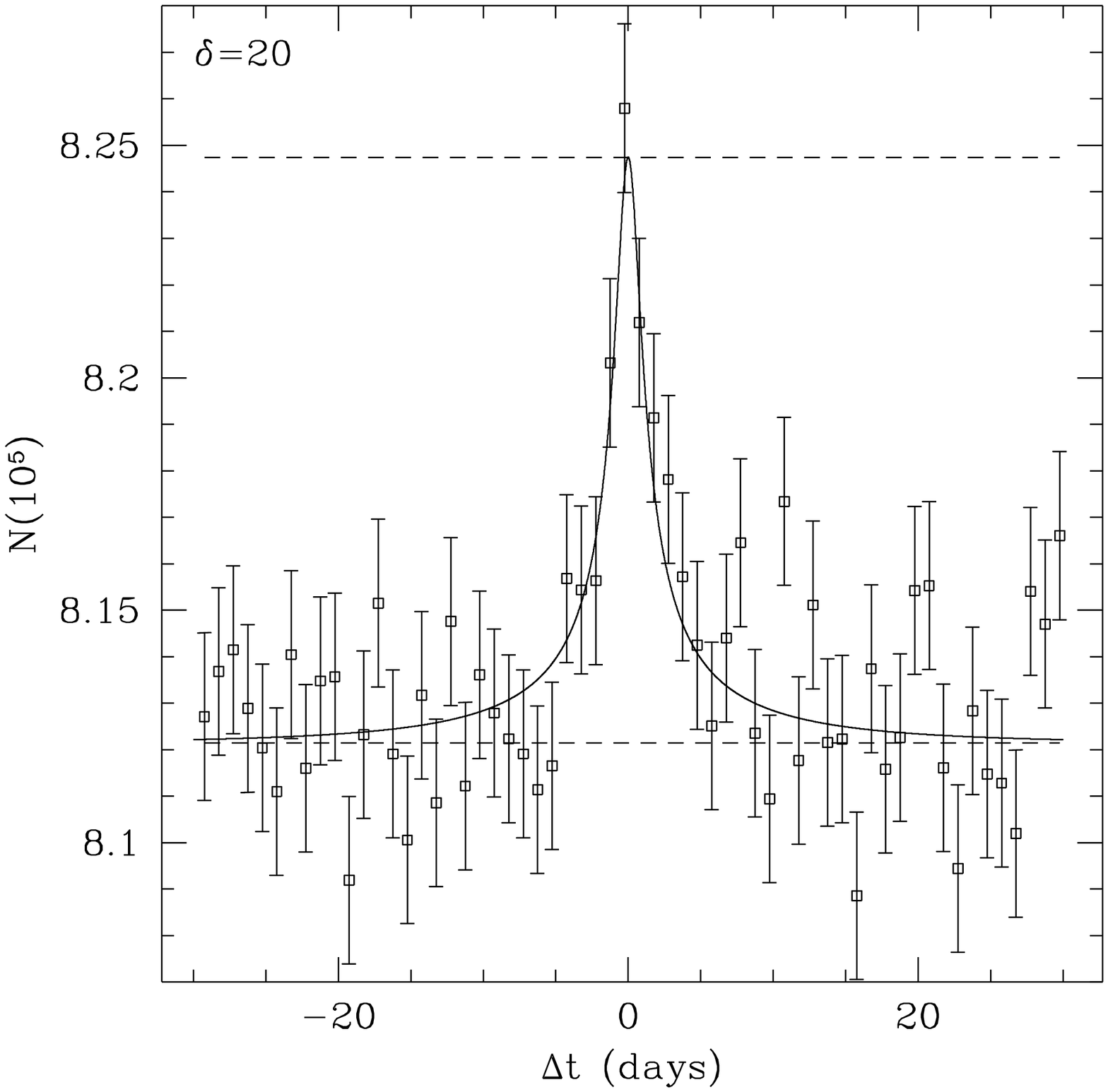}
\epsfig{width=3.2in,file=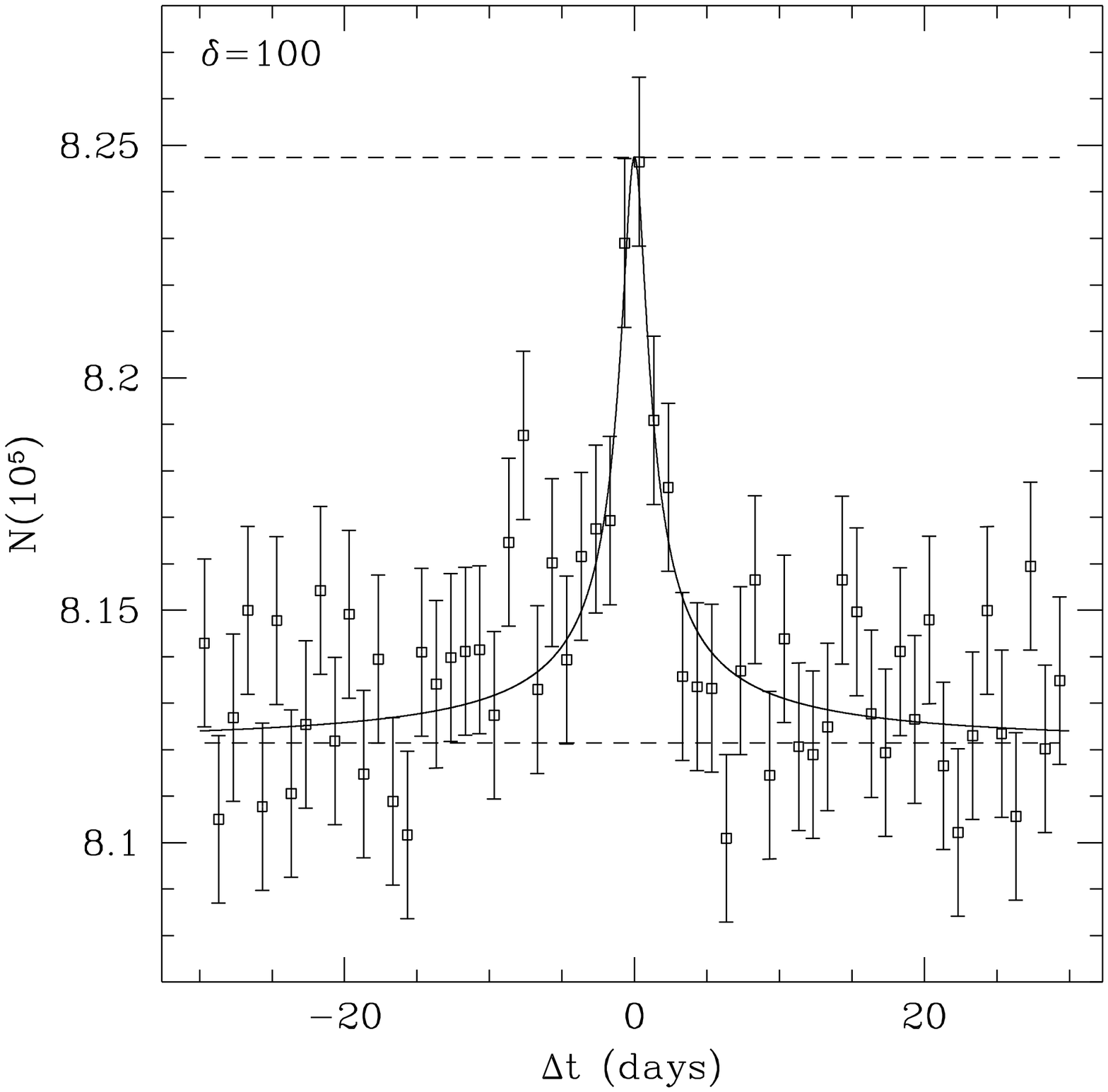}\\
\epsfig{width=3.2in,file=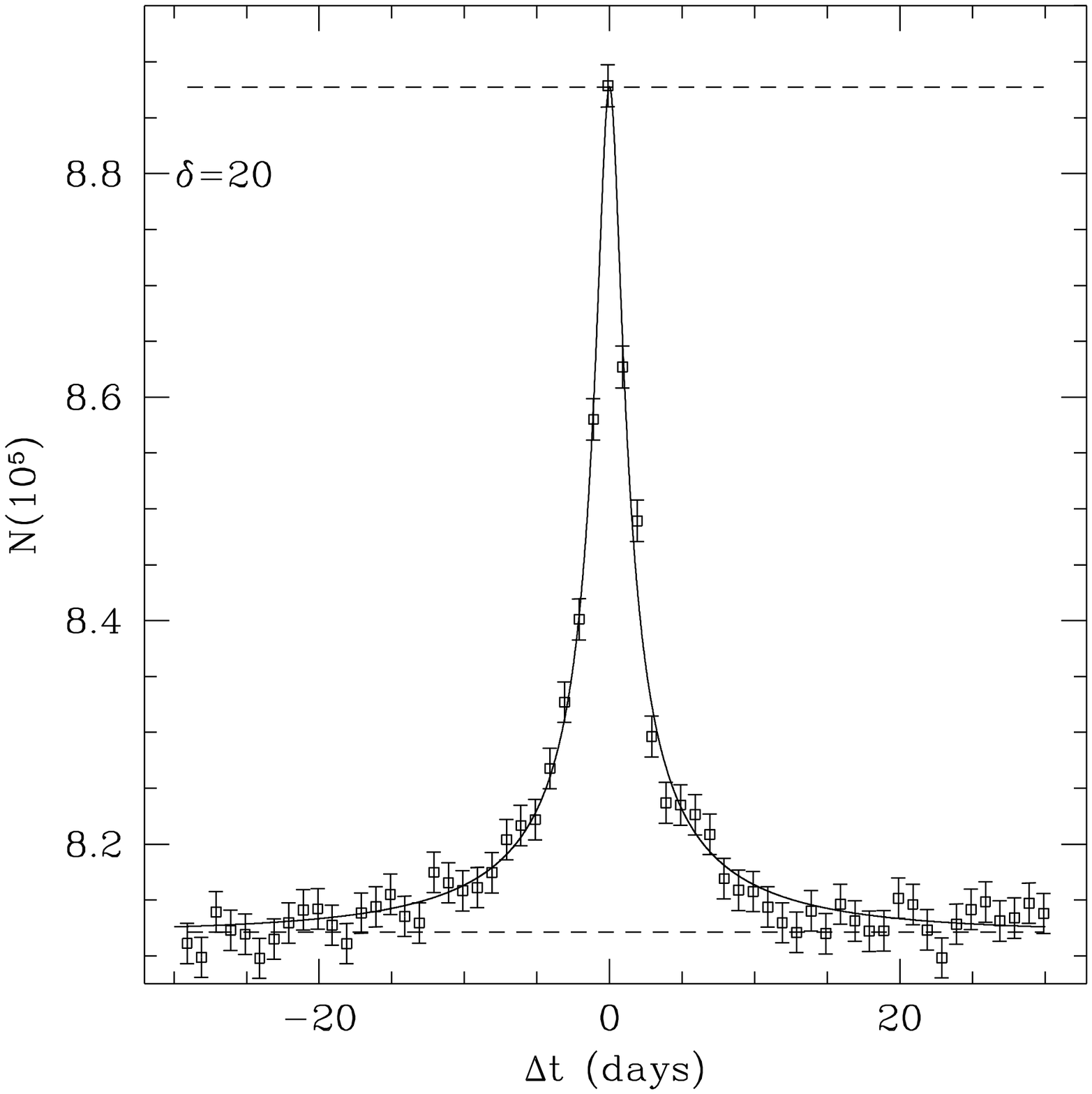}
\epsfig{width=3.2in,file=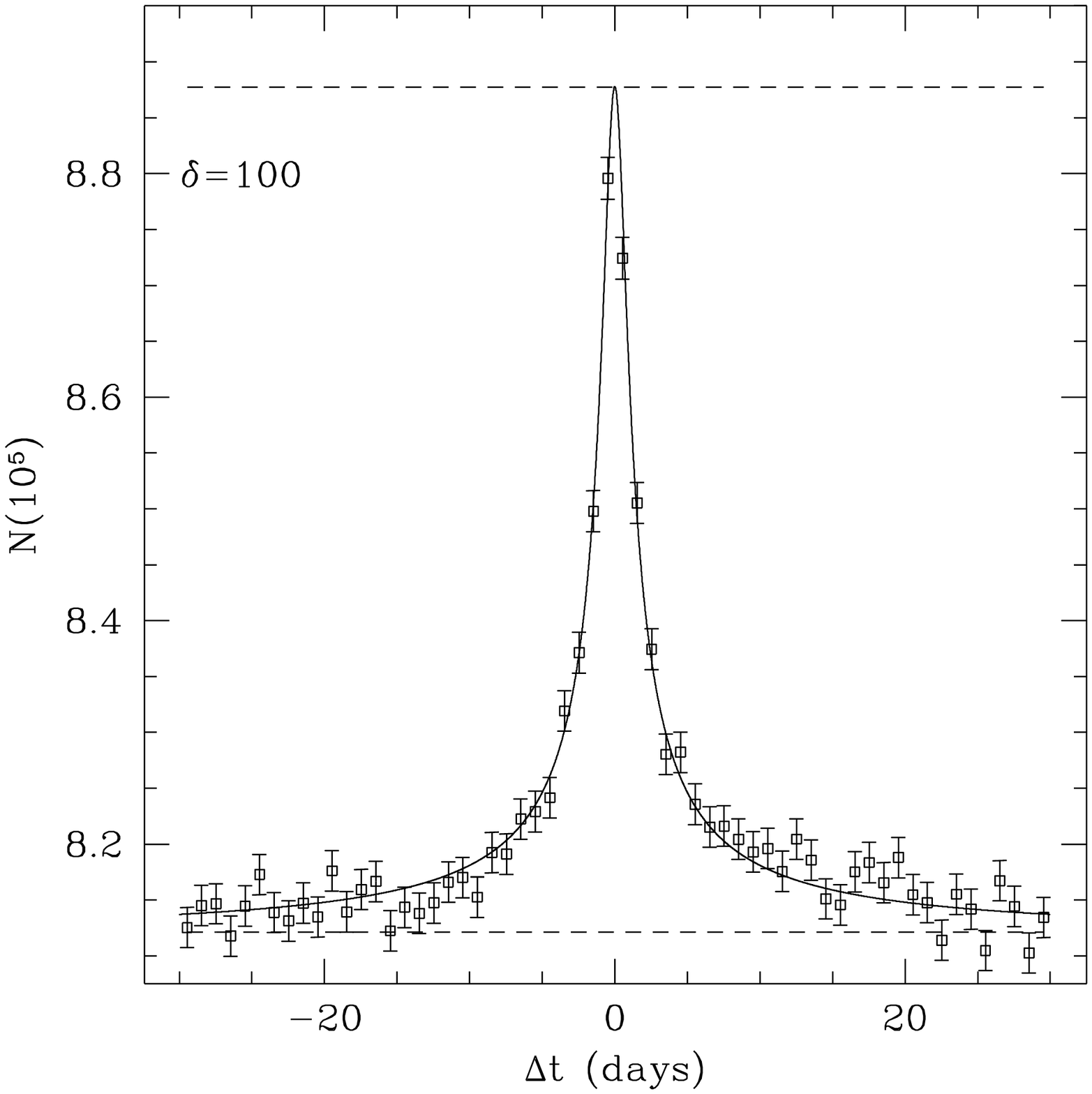}

\newpage
\begin{figure}
\caption{Histograms of {\em measured} $t_{\sigma 10}/\tf$.  In all cases,
$\tf=3$ days, and the peak signal to noise is $Q_0=42$.  The five dashed
vertical lines correspond, from left to right, to the true values when
$\delta=10,20,50,100,\infty$.  The four points and error bars indicate the
mean measured values and errors.}
\label{fig:hist}
\end{figure}
\noindent
\epsfig{width=3.2in,file=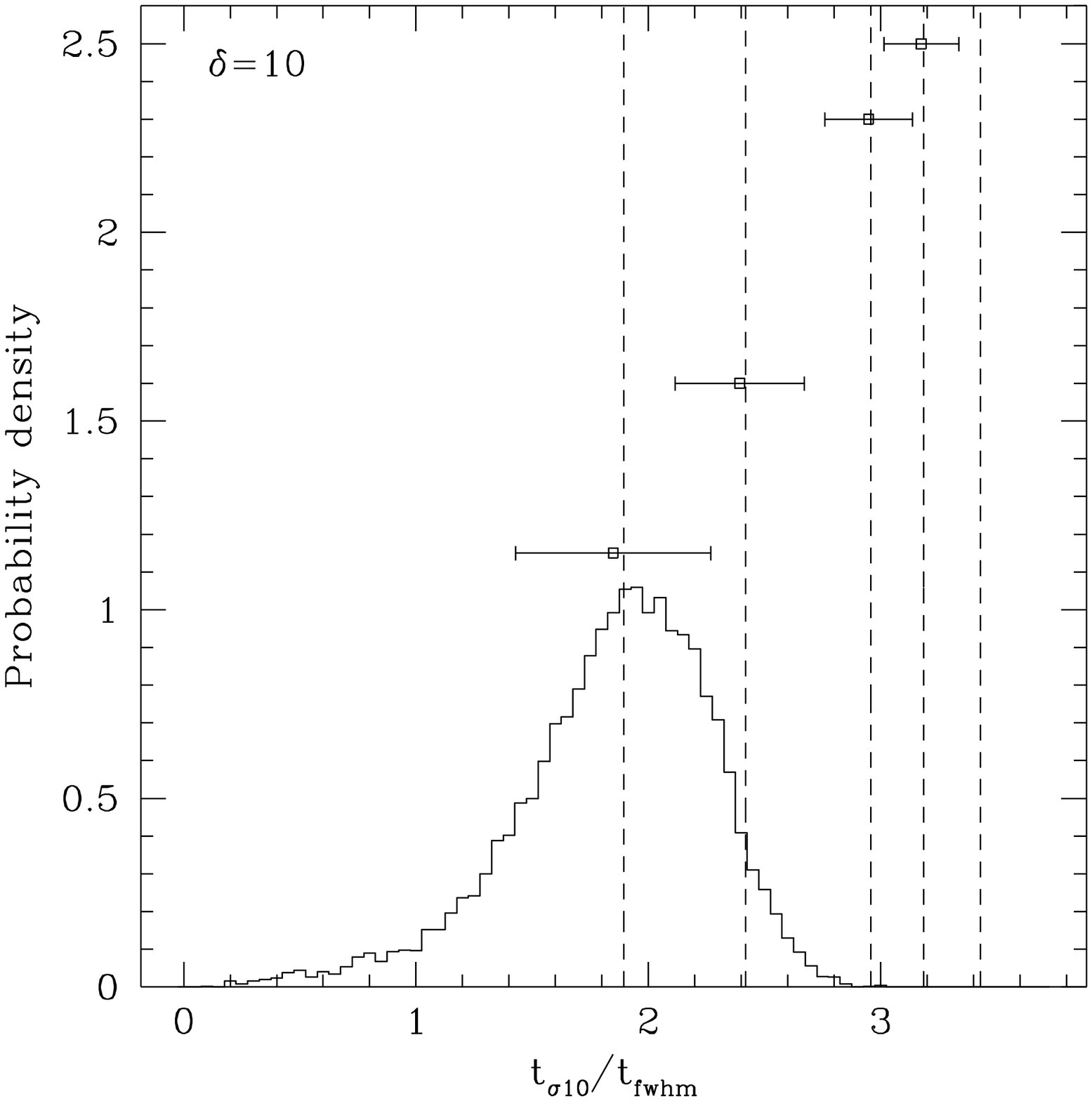}
\epsfig{width=3.2in,file=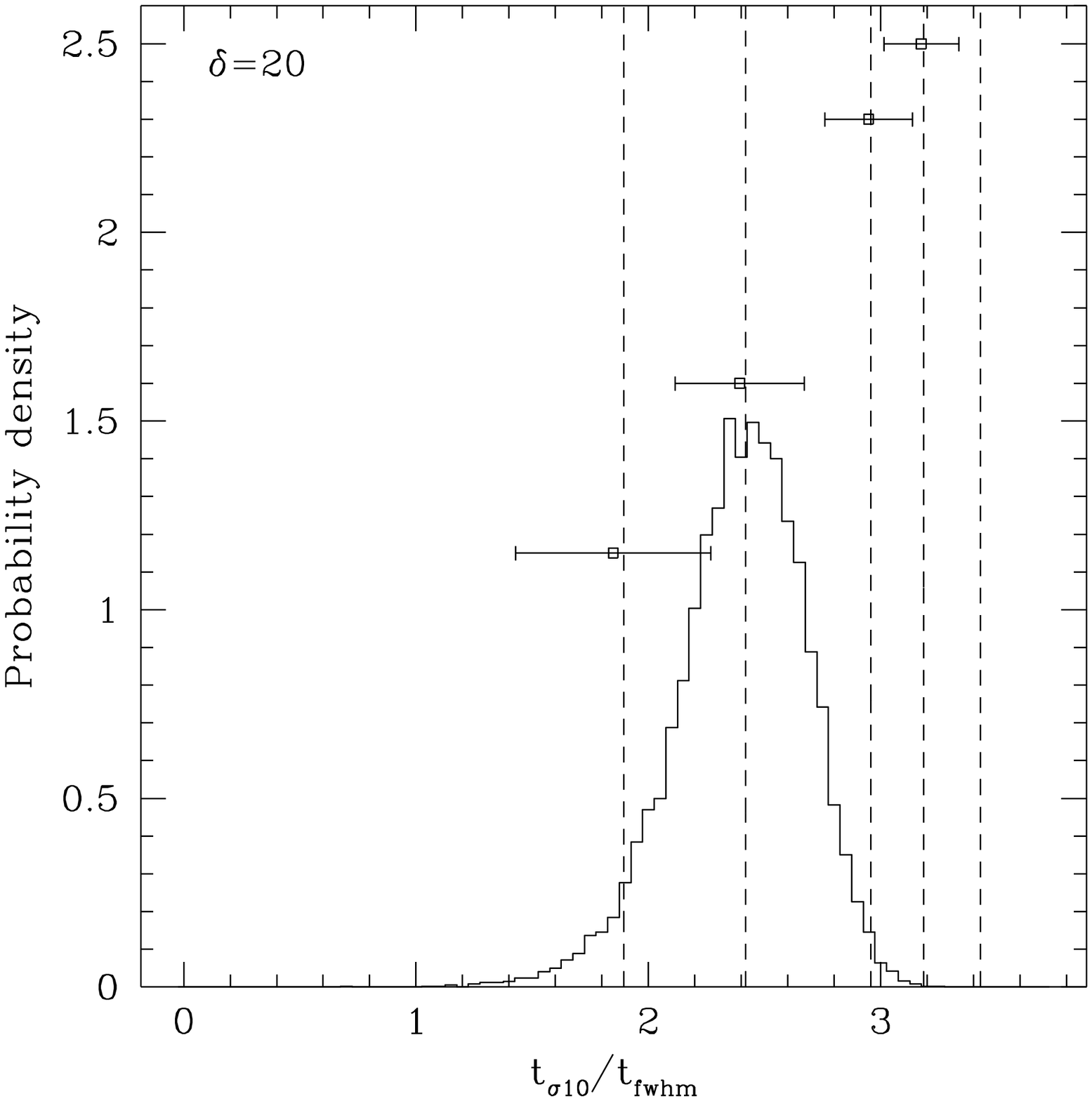}\\
\epsfig{width=3.2in,file=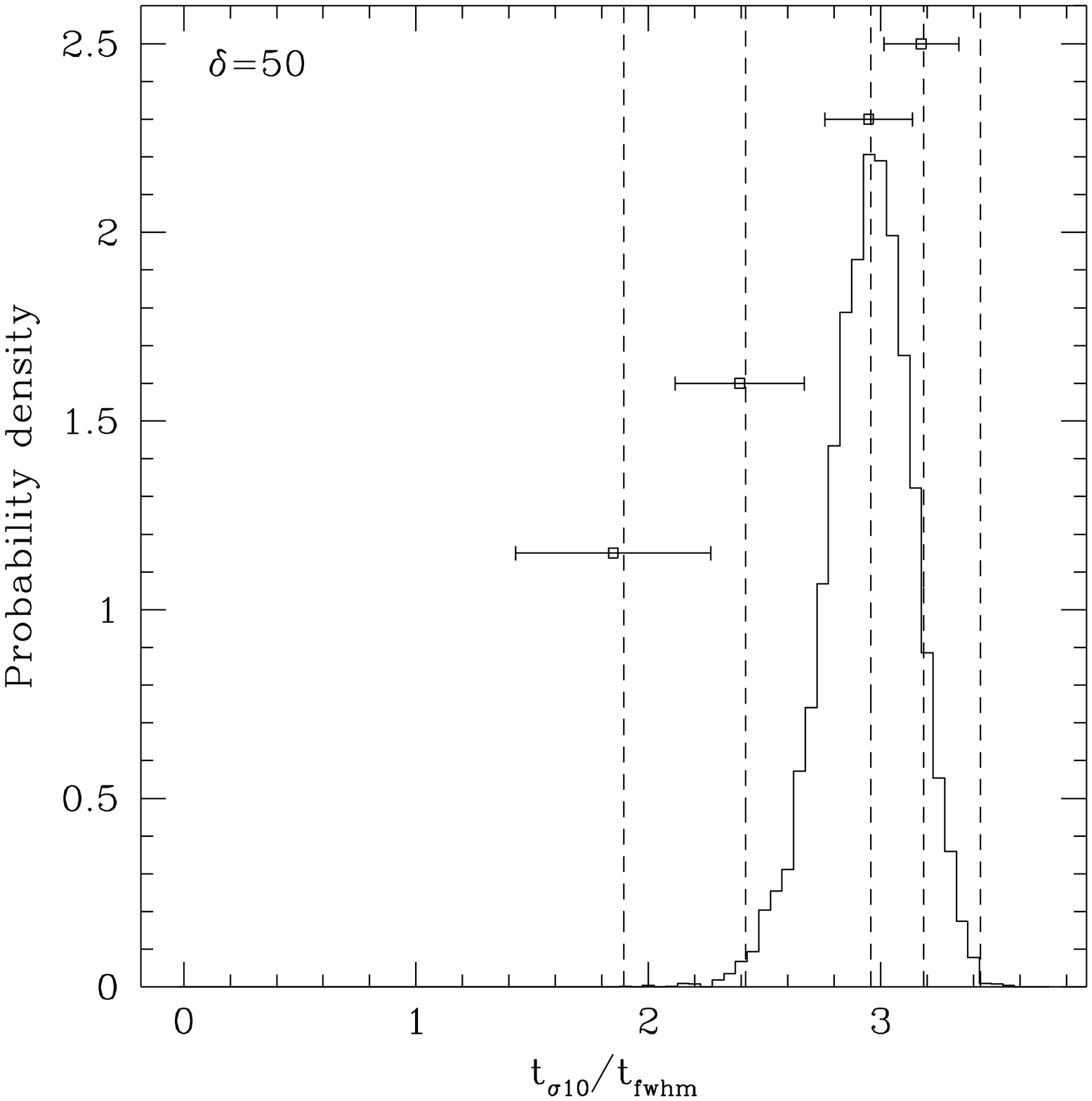}
\epsfig{width=3.2in,file=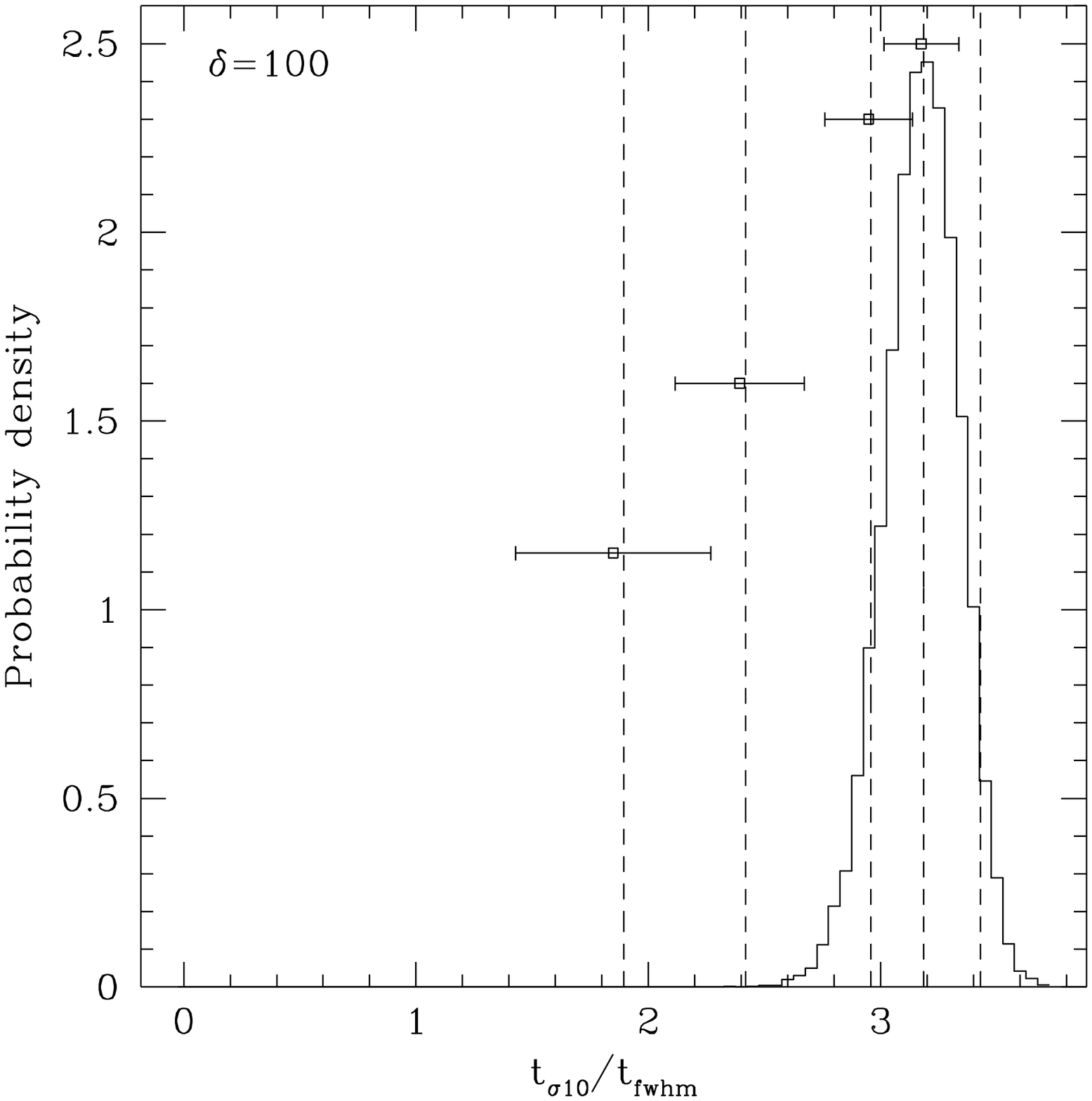}

\end{document}